\documentclass[twocolumn]{aastex61}
\pdfoutput=1
\usepackage{graphicx}
\usepackage{amsmath}
\usepackage{epsfig}
\usepackage{bm}

\newcommand \kms {km s$^{-1} \ $}

\shorttitle{STELLAR WIND \& SNR}
\shortauthors{M. F. Zhang et al.}
\begin{document}
\title{How does the stellar wind influence the radio morphology of a supernova remnant?}

\author{M. F. Zhang}
\affiliation{Key Laboratory of Optical Astronomy, National Astronomical Observatories, Chinese Academy of Sciences,
Beijing 100012, China}
\affiliation{University of Chinese Academy of Sciences, 19A Yuquan Road, Shijingshan District, Beijing 100049, China}

\author{W. W. Tian}
\affiliation{Key Laboratory of Optical Astronomy, National Astronomical Observatories, Chinese Academy of Sciences,
Beijing 100012, China}
\affiliation{University of Chinese Academy of Sciences, 19A Yuquan Road, Shijingshan District, Beijing 100049, China}

\author{D. Wu}
\affiliation{Key Laboratory of Optical Astronomy, National Astronomical Observatories, Chinese Academy of Sciences,
Beijing 100012, China}
\affiliation{University of Chinese Academy of Sciences, 19A Yuquan Road, Shijingshan District, Beijing 100049, China}

%
%

\correspondingauthor{M. F. Zhang}
\email{zmf@bao.ac.cn}

\label{firstpage}
\begin{abstract}
We simulate the evolutions of the stellar wind and the supernova remnant (SNR) originating from a runaway massive star
in an uniform Galactic environment based on the three-dimensional magnetohydrodynamics models.
Taking the stellar wind into consideration, we can explain the radio morphologies of many supernova remnants.
The directions of the kinematic velocity of the progenitor, the magnetic field and the line of sight are the most important
factors influencing the morphologies.
If the velocity is perpendicular to the magnetic field, the simulation will give us two different unilateral SNRs and a
bilateral symmetric SNR.
If the velocity is parallel to the magnetic field, we can obtain a bilateral asymmetric SNR and a quasi-circular
SNR.
Our simulations show the stellar wind plays a key role in the radio evolution of a SNR, which implies the
Galactic global density and magnetic field distribution play a secondary role in shaping a SNR.
\end{abstract}


\section{Introduction}
A massive star dies, then forms a supernova remnant (SNR).
This process produces heavy elements, dusts and cosmic rays, which has important impact on the Galactic interstellar
medium (ISM).
To understand this process, we need study the evolution of SNRs.
\citet{Truelove1999} and \citet{Cioffi1988a} did many analytical and numerical calculations about the evolution.
Comparing the results with the observations, they developed a practical model.
However, there is usually the diverse surrounding environment which will influence the evolution of SNRs.
As a result, the radio morphologies of SNRs are various.
The practical model can explain some regular morphologies, such as bilateral symmetric and circular SNRs, but is powerless to
explain more complex morphologies.
These morphologies can help us infer some important natures of SNRs, so it is significant to study them in detail.

The numerical simulation is an effective method to describe the surrounding environment and obtain the evolution images of
a SNR at different phases.
With the improvement of the computation ability, the two-dimensional (2D) hydrodynamics (HD) simulation shows its power in studying
the magnetic amplification, the diffusive shock acceleration and the instability of SNRs \citep{Jun1996,Kang2006,Fang2012}.
Recently, we can perform three-dimensional (3D) simulations, and also convert the simulation results to radio,
optical or X-ray images in order to compare with observations \citep{Orlando2007,Meyer2015,Zhang2017}.
\citet{Orlando2007} tried to explain asymmetric morphologies of some bilateral supernova remnants by assuming inhomogeneous
density and magnetic field.
They simulated some asymmetric structures in SNRs, but did not describe how the assuming surrounding environment
is formed around the SNRs.
\citet{West2016} thought the surrounding environments are mainly influenced by the Galactic global ISM distribution and
applied a method of magnetohydrodynamics (MHD) simulation to study the Galactic magnetic field model.
They partly explains the assumed surrounding environments by \citet{Orlando2007}, but cannot well simulate many asymmetric
structures.
Thus, there should probably be another factor influencing the surrounding environments.

This factor is possibly the stellar wind of the progenitor.
The progenitor runs in the ISM and blows a stellar wind bubble, which leads to inhomogeneous density distribution and magnetic
field structure.
This certainly influences the following remnant 's evolution and its radio morphology when a supernova explodes in such a bubble.
This assumption is self-consistent and supported by theoretical calculations and observations \citep{Chen1995,Zhang1996,Foster2004,Lee2010}.
\citet{Meyer2015} simulated the stellar wind, then took the result as the initial condition of the SNR simulation.
They concluded that the stellar wind will strongly shape the density distribution of the SNRs.
They only performed the 2D HD simulations and did not obtain the radio images.
The crucial parameters of the 3D MHD simulation include the density and the magnetic field of the ISM, the spatial velocity
and the stellar wind of the progenitor, the explosion energy and the mass of the supernova.
It is impossible to test all combinations of these parameters by now.
In particular, there are two vectorial parameters, the magnetic field of the ISM and the velocity of the progenitor.
Each vector has three components, which largely complicates the conditions that one has to take into account for the 3D simulation.

We in the paper present a 3D MHD simulation where these parameters are fixed but the relative directions of the magnetic
field and the velocity of the progenitor.
We perform two simulations, one for the magnetic field in perpendicular to the velocity, one for the magnetic field in parallel
to the velocity.
In the following text, we call the former the perpendicular simulation and the latter the parallel simulation.
Using canonical values of a massive star, we may obtain many radio morphologies of SNRs based on such a simplification.
We also count different types of SNRs, so that we can better understand our simulation results.

In Sect.2, we describe the simulation model and list the parameters we use.
In Sect.3, we present and discuss the results.
Sect.4 is a summary.

\section{simulation model}
The simulation model is based on a 3D MHD frame with a grid of 128 $\times$ 128 $\times$ 128.
The spatial scale is set to 60 pc $\times$ 60 pc $\times$ 60 pc, i.e. its resolution is 0.47 pc pixel$^{-1}$.
The viscosity and the gravitation have little influence on the simulation, so we ignore them.
The cooling and heating effect mainly influences the luminosity of optical and X-ray radiation,
and we mainly focus radio radiation, so they are not included in the simulation.
In the stellar wind simulation, the thermal conduction is an important process \citep{Meyer2014}, which can
govern the shape, the size and the structure of the stellar winds.
However, it is not the dominant factor in the SNR simulation, so we only discuss its influence in the perpendicular
simulation.
The simulation is based on the ideal conservation equation set:
\begin{equation}
    \begin{cases}
      \dfrac{\partial \rho}{\partial t} + \nabla \cdot (\rho \bm{v}) = 0 , \\
      \dfrac{\partial \rho \bm{v}}{\partial t} + \nabla \cdot (\rho \bm{vv} - \bm{BB}) + \nabla P^* = 0 , \\
      \dfrac{\partial E}{\partial t} + \nabla \cdot [(E+P^*)\bm{v} - \bm{B}(\bm{v} \cdot \bm{B})] = 0 , \\
      \dfrac{\partial \bm{B}}{\partial t} + \nabla \times (\bm{v} \times \bm{B}) = 0,
    \end{cases}
\end{equation}
in which, $\rho$ is mass density, $\bm{v}$ is velocity, $\bm{B}$ is magnetic field intensity, $P^*$ is total
pressure, and $E$ is total energy density.

The simulation contains two models, the stellar wind model and the supernova remnant model.
At first, we simulate the evolution of the stellar wind, and the results are taken as the initial conditions in the
SNR simulation.
Then we perform the SNR simulation and convert the results to relative radio flux density images.
Finally, we compare the simulation radio images with the observed radio images.

We perform the simulations using a code, PLUTO \footnote{http://plutocode.ph.unito.it/}
\citep{Mignone2007, Mignone2012},
and summary the parameters in Table.~\ref{table:parameters}.
The parameters that we do not show the references are just the canonical values we estimate.

\begin{table}
  \caption{Summary of Simulation Parameters}
  \label{table:parameters}
  \centering
  \begin{tabular}{l l l}
      \hline\hline
      Parameters                      & Value            & References               \\
      \hline
      Stellar Wind Parameters\\
      \hline
      Progenitor Velocity             & 40 \kms   & 1 \\
      Mass-Loss Rate                  & 3 $\times$ 10$^{-6}$ M$_{\odot}$ yr$^{-1}$ & 2 \\
      Stellar Wind Velocity           & 800 \kms  & 2 \\
      Stellar Wind Density            & 0.05\ cm$^{-3}$  & 2 \\
      Inner Radius                    & 0.5 pc \\
      Evolution Time                  & 1 million years  & 1 \\
      \hline
      SNR Parameters\\
      \hline
      Ejecta Mass                     & 15.3 M$_{\odot}$ & 3\\
      Initial Explosion Energy        & 1.3$\times$ 10$^{51}$ ergs & 4, 5\\
      Initial Radius                  & 4 pc             &\\
      Initial Time                    & 650 years        & 6\\
      \hline
      Other Parameters\\
      \hline
      Mean Density                    & 0.5\ cm$^{-3}$   & 7, 8\\
      Magnetic Field Intensity        & 9 $\mu$G         & 9\\
      Mean Atomic Weight              & 1.3              &\\
      Adiabatic Coefficient           & 1.7              &\\
      Synchrotron Index ($\beta$)     & 0.5              &\\
      \hline
  \end{tabular}\\
  \tablerefs{(1)\citealt{Meyer2014}; (2)\citealt{Meyer2015}; (3)\citealt{Sukhbold2016};
  (4)\citealt{Poznanski2013}; (5)\citealt{Mueller2016a}; (6)\citealt{Leahy2017a}; (7)\citealt{Nakanishi2006};
  (8)\citealt{Nakanishi2016}; (9)\citealt{Haverkorn2015}}
\end{table}

\subsection{The stellar wind model}
How the stellar winds of runaway massive stars evolve is still an unsolved problem, so we only use a reasonable simplified model.
If the stellar winds can influence the SNRs obviously, their spatial scales should be similar to SNRs.
The typical diameters of SNRs are usually several parsecs (pcs).
\citet{Meyer2014} showed that the mass of the star should be at least 40 M$_{\odot}$ to reach such a scale, if the speed of
the star is 40 \kms.
Lower mass means lower speed \citep{Mackey2015}, but lower speed means lower asymmetry, which is inconsistent with the
aim of this paper.
We therefore choose the mass 40 M$_{\odot}$ and the speed 40 \kms as the initial parameters in our simulation.
It is known that the star's life is composed of the main sequence (MS) and the red supergiant (RSG) phase.
However, our tests show the stellar wind in main sequence phase has little impact on the evolution of a SNR, so we only simulate
it for the last one million years.

The mass loss of a 40 M$_{\odot}$ star usually varies from 1 $\times$ 10$^{-6}$ to 1 $\times$ 10$^{-5}$ M$_{\odot}$ yr$^{-1}$
during the last one million years of the star's life \citep{Meyer2014, vanMarle2012, vanMarle2015}, so we use a mass-loss rate of
3 $\times$ 10$^{-6}$ M$_{\odot}$ yr$^{-1}$ for simplicity.
Here we warn readers that it is not reality to accurately estimate the mass-loss rate of a massive star so far
\citep{Meyer2014a, Gvaramadze2014}.
Also, we set the inner radius as 0.5 pc, i.e. the stellar wind is generated from such a small region in the simulation.
This radius is large enough to guarantee the wind blows spherically in the square grid of numerical simulation and small enough to be
consistent with the simplified stellar wind model.
The mass-loss rate \textit{$\dot{M}$}, the inner radius \textit{r}, the velocity \textit{v} and the mass density \textit{$\rho$} of
the stellar wind are linked by

\begin{equation}
  \begin{aligned}
    \dot{M}=4\pi r^2\rho v.
  \end{aligned}
\end{equation}

The initial velocity of the stellar wind originating from the progenitor will not change in 0.5 pc, if we assume it propagates
freely in such a short radius.
Then the velocity should be about 800 \kms and the density is about 0.05 cm$^{-3}$ \citep{Meyer2014}.

In addition, we set the initial surrounding environment before the stellar wind evolution.
We assume the ISM is ideal gas, where the mean atomic weight is 1.3 and the adiabatic coefficient is 1.7.
We set a uniform magnetic field of 9 $\mu$G \citep{Haverkorn2015} and a uniform ISM number density of 0.5 cm$^{-3}$
\citep{Nakanishi2006,Nakanishi2016}, the typical values of the Galactic ISM.
The environment is usually inhomogeneous, which will result in a more complex radio morphology in the simulation.
However, we only want to test how the SNRs are influenced by the stellar winds, so we use a homogeneous ISM in this work.

\subsection{The supernova remnant model}
The evolution of a SNR is divided into three phases, the ejecta-dominated (ED) phase, the Sedov-Taylor (ST) phase and
the pressure-driven snowplow (PDS) phase \citep{Truelove1999}.
The first two phases are classified as "nonradiative", but the radiative loss becomes important in the PDS phase.
Our simulations only cover the first two phases, so we do not need to estimate radiative loss.
For a 40 M$_{\odot}$ star, the ejecta mass is about 15.3 M$_{\odot}$ \citep{Sukhbold2016} and the explosion energy is
about 3.6 $\times$ 10$^{51}$ erg according to the function \citep{Poznanski2013,Mueller2016a},

\begin{equation}
  \begin{aligned}
    log(E/10^{50}erg)=2.09log(M_{ej}/M_{\odot})-1.78.
  \end{aligned}
\end{equation}

To simulate a spherically symmetric explosion, we set an initial radius as 4 pc.
The shock wave of the supernova explosion will spend 650 years to reach 4 pc.
Because the ST phase starts from 1365 years for such a star \citep{Leahy2017a}, it is still in ED phase.
Therefore, we can obtain the 650-years evolution directly from the existed theory \citep{Truelove1999}
which gives the density, pressure and velocity profile.
The magnetic field is not important at this time, so we ignore it here.
In short, the initial conditions are the evolution results after 650 years.

Next we start to simulate the evolution of a SNR in the surrounding environment blown by the stellar wind.
Our simulation has shown the density, the magnetic field, the velocity and the pressure in the whole simulation space.
We further convert these simulation results into radio images in order to compare with real observations.

Assuming the radio emission is totally from synchrotron mechanism, we obtain the radio flux volume density by employing
$i(\nu)=C\rho B_{\perp}^{\beta + 1}\nu^{-\beta}$ \citep{Orlando2007}, in which $\nu$ is the radiation frequency,
C a constant, $\rho$ the density, $B_{\perp}$ the magnetic field perpendicular to the line of sight (LoS) and
$\beta$ the synchrotron spectral index.
The absolute radio flux density is dependent on the constant C, but C contains electron acceleration efficiency which is
difficult to be obtained.
Moreover, the $\nu^{-\beta}$ is also excluded from the equation, because it is meaningless if we do not want to
calculate the absolute radio flux density.
As a result, the final equation used in this work is $i(\nu)=\rho B_{\perp}^{\beta + 1}$.
Then we integrate the $i(\nu)$ along the LoS to obtain relative radio flux density.
The resolution of the simulation is usually higher than the observation, so we smooth the simulation radio images by using
a 2D Gaussian function with $\sigma = 1$.
\begin{figure*}
    \centering
    \includegraphics[width=0.325\textwidth]{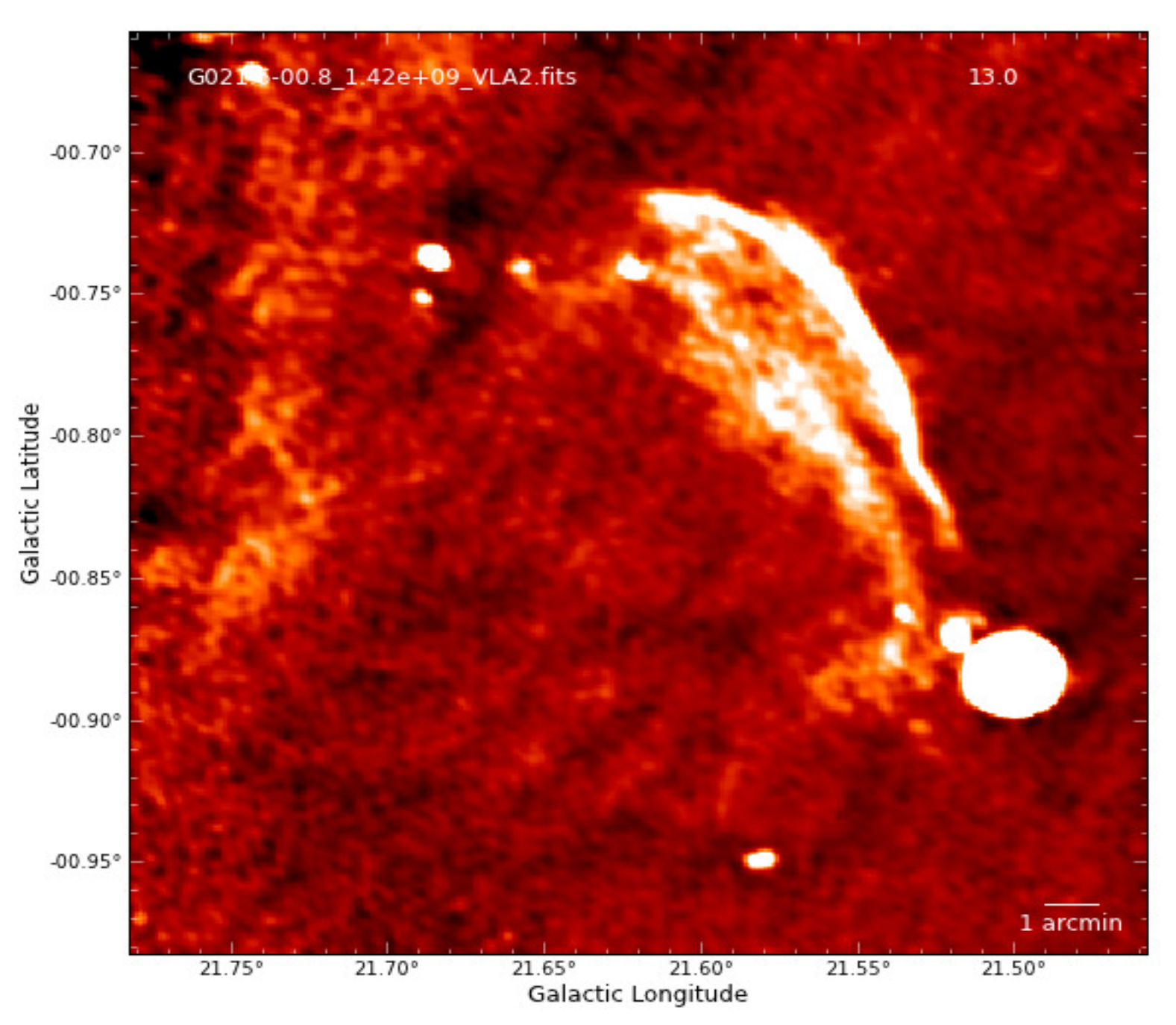}
    \includegraphics[width=0.325\textwidth]{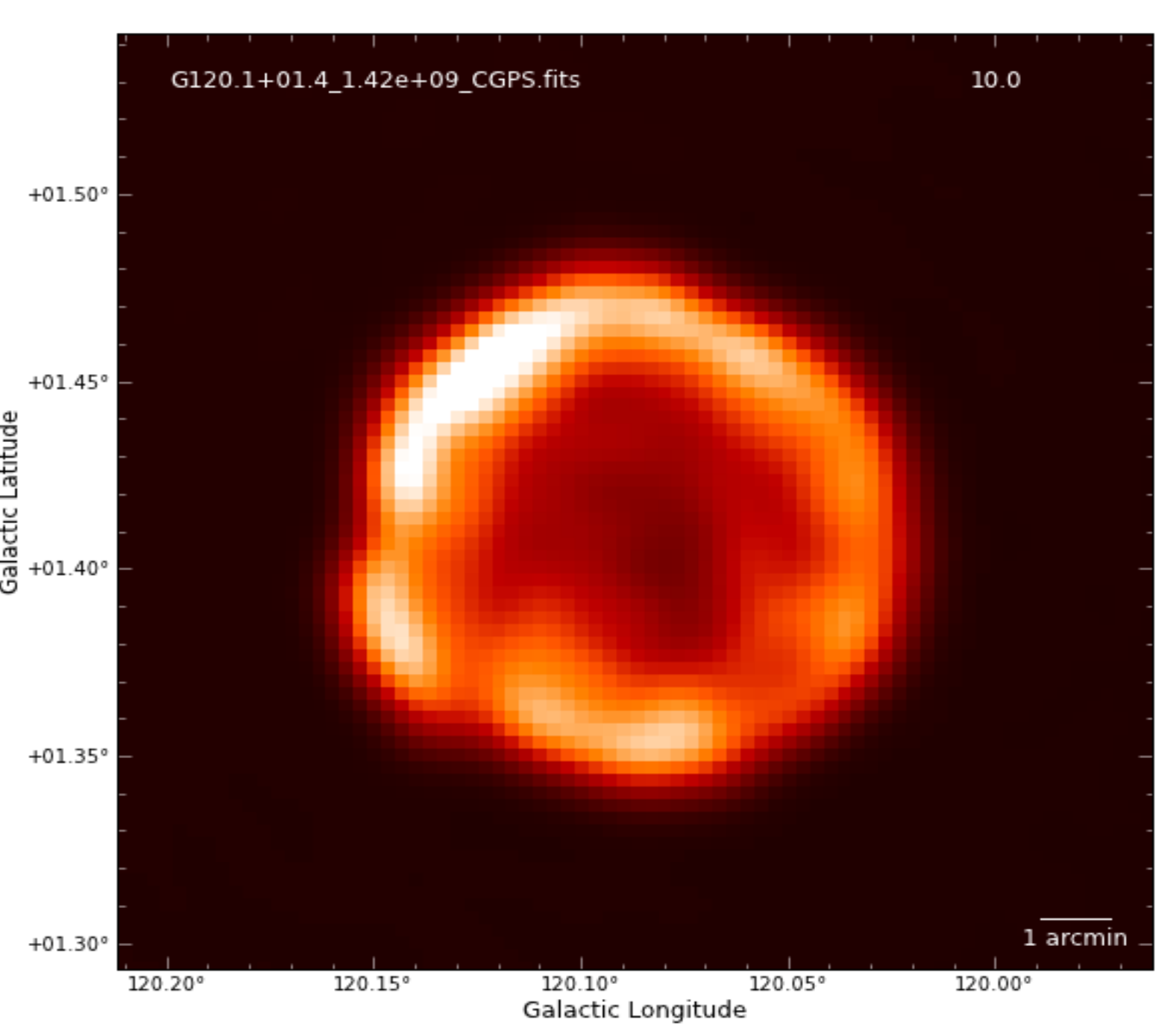}
    \includegraphics[width=0.325\textwidth]{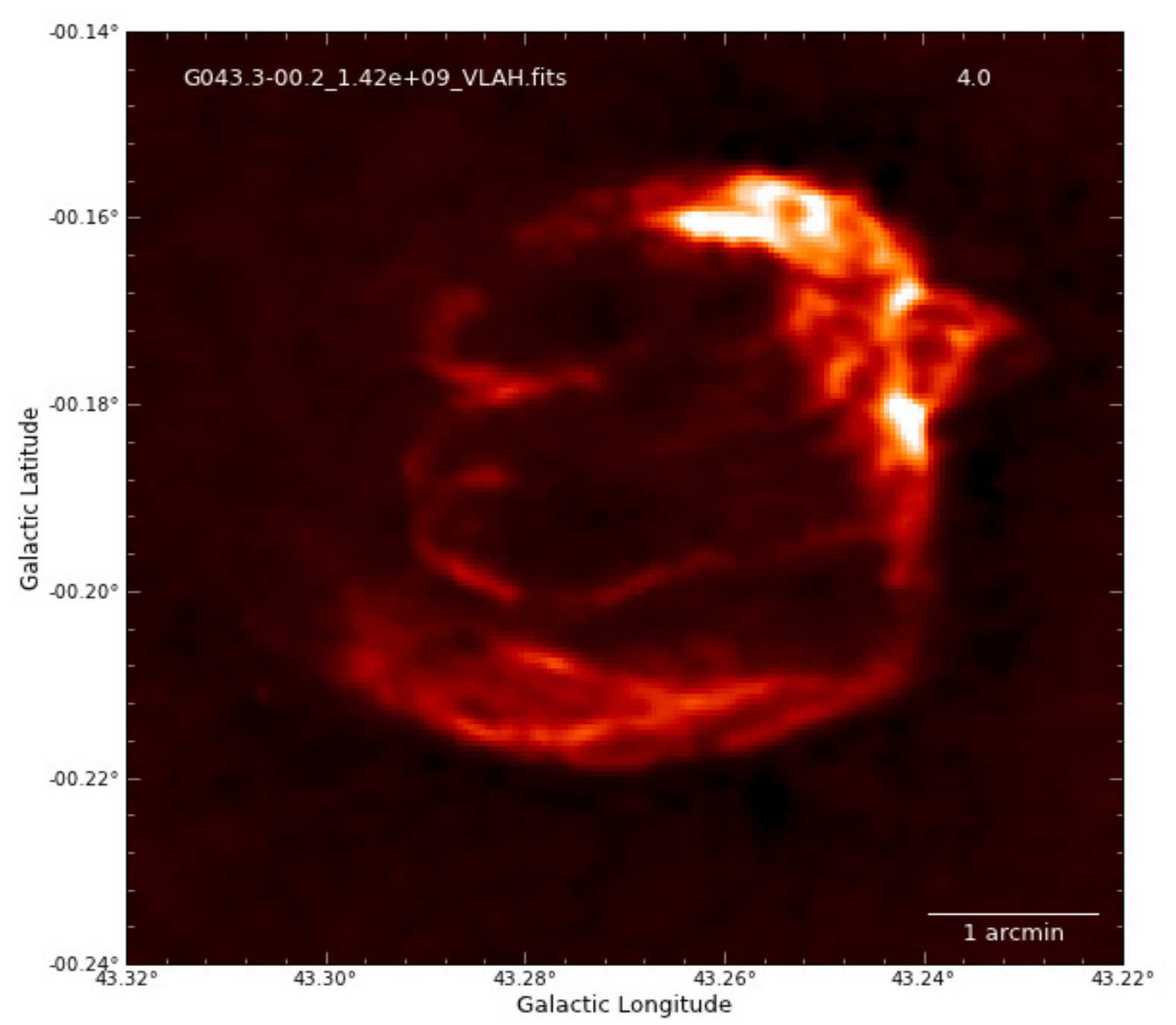}
    \caption{The typical multi-layers, circular and irregular SNRs: G21.6-0.8, G120.1+1.4 and G43.3-0.2, respectively. }
\label{fig:stat}
\end{figure*}

\begin{table*}
  \caption{Statistics of different SNRs}
  \label{table:stat}
  \centering
  \begin{tabular}{l l l}
      \hline\hline
      Types                           & Numbers           & Samples               \\
      \hline
      unilateral small-radian         & 35                &G4.2-3.5, G5.9+3.1, G6.1+0.5, G6.4+4.0, G7.0-0.1, G7.2+0.2,
      G11.1+0.1, G11.1-0.7, \\& & G12.2+0.3,  G14.3+0.1, G17.4-0.1, G24.7-0.6, G49.2-0.7, G57.2+0.8, G59.8+1.2, \\& & G65.1+0.6,
      G310.8-0.4, G327.4+1.0, G338.1+0.4, G348.5-0.0, \\& & G348.7+0.3, G350.0-2.0, G351.7+0.8, G351.9-0.9,  G354.1+0.1,
      G359.0-0.9\\
      \hline
      unilateral large-radian         & 15                &G0.0+0.0, G1.9+0.3, G3.8+0.3, G8.3-0.0, G9.8+0.6, G18.6-0.2,
      G18.8+0.3, G33.2-0.6,  \\& & G55.7+3.4, G66.0-0.0, G116.9+0.2, G119.5+10.2, G298.6-0.0, G321.9-1.1, G342.1+0.9\\
      \hline
      bilateral symmetric             & 17                &G0.9+0.1, G1.0-0.1, G3.7-0.2, G8.7-5.0, G16.2-2.7, G21.0-0.4,
      G23.3-0.3, G36.6+2.6, \\& & G59.5+0.1,  G65.3+5.7, G296.5+10.0, G321.9-0.3, G327.6+14.6, G332.0+0.2, G349.2-0.1,
      \\& & G353.9-2.0, G356.3-1.5\\
      \hline
      bilateral asymmetric            & 11                &G11.0-0.0, G21.8-0.6, G29.7-0.3, G42.8+0.6, G53.6-2.2, G54.4-0.3,
      G64.5+0.9,  \\& & G304.6+0.1, G348.5+0.1, G350.1-0.3, G352.7-0.1\\
      \hline
      multi-layers                     & 13                &G21.6-0.8, G24.7+0.6, G46.8-0.3, G85.4+0.7, G93.3+6.9, G109.1-1.0,
      G284.3-1.8,  \\& & G286.5-1.2,  G318.9+0.4, G320.6-1.6, G327.4+0.4, G358.1+1.0, G358.5-0.9\\
      \hline
      circular                        & 42                &G4.5+6.8, G5.2-2.6, G6.5-0.4, G11.2-0.3, G11.4-0.1, G15.9+0.2,
      G16.7+0.1, G18.1-0.1, \\& & G21.5-0.9,  G27.4+0.0, G69.7+1.0, G82.2+5.3, G83.0-0.3, G84.2-0.8, G111.7-2.1,\\& & G120.1+1.4,
      G132.7+1.3,  G179.0+2.6, G180.0-1.7, G184.6-5.8, G261.9+5.5, G290.1-0.8, \\& & G299.2-2.9, G301.4-1.0, G302.3+0.7,
      G308.1-0.7, G310.6-0.3, G311.5-0.3, G315.4-2.3, \\& & G322.5-0.1, G326.3-1.8, G327.1-1.1, G327.2-0.1,  G332.4-0.4,
      G337.3+1.0, G346.6-0.2, \\& & G354.8-0.8, G355.6-0.0, G355.9-2.5, G356.2+4.5, G358.0+3.8, G359.1-0.5\\
      \hline
      irregular                       & 155               &\\
      \hline
  \end{tabular}\\
\end{table*}
\section{Results and Discussion}

We show the results and compare them with the observations in this section.

Based on \citet{West2016}'s collection of all radio SNRs' images, we classify the SNRs to seven types: unilateral small-radian,
unilateral large-radian, bilateral symmetric, bilateral asymmetric, multi-layers, circular and irregular.
A multi-layers SNR means there are two or more layers on one or two sides.
The typical multi-layers, circular and irregular SNRs are shown in Figure~\ref{fig:stat}.
The statistics of the seven types is listed in Table.~\ref{table:stat}.
We only select 288 SNRs in this statistics, because other images are obscure.
However, we list all samples except for the irregular type for the convenience of readers.

\subsection{Perpendicular Simulation}
\begin{figure*}
    \centering
    \includegraphics[width=0.325\textwidth]{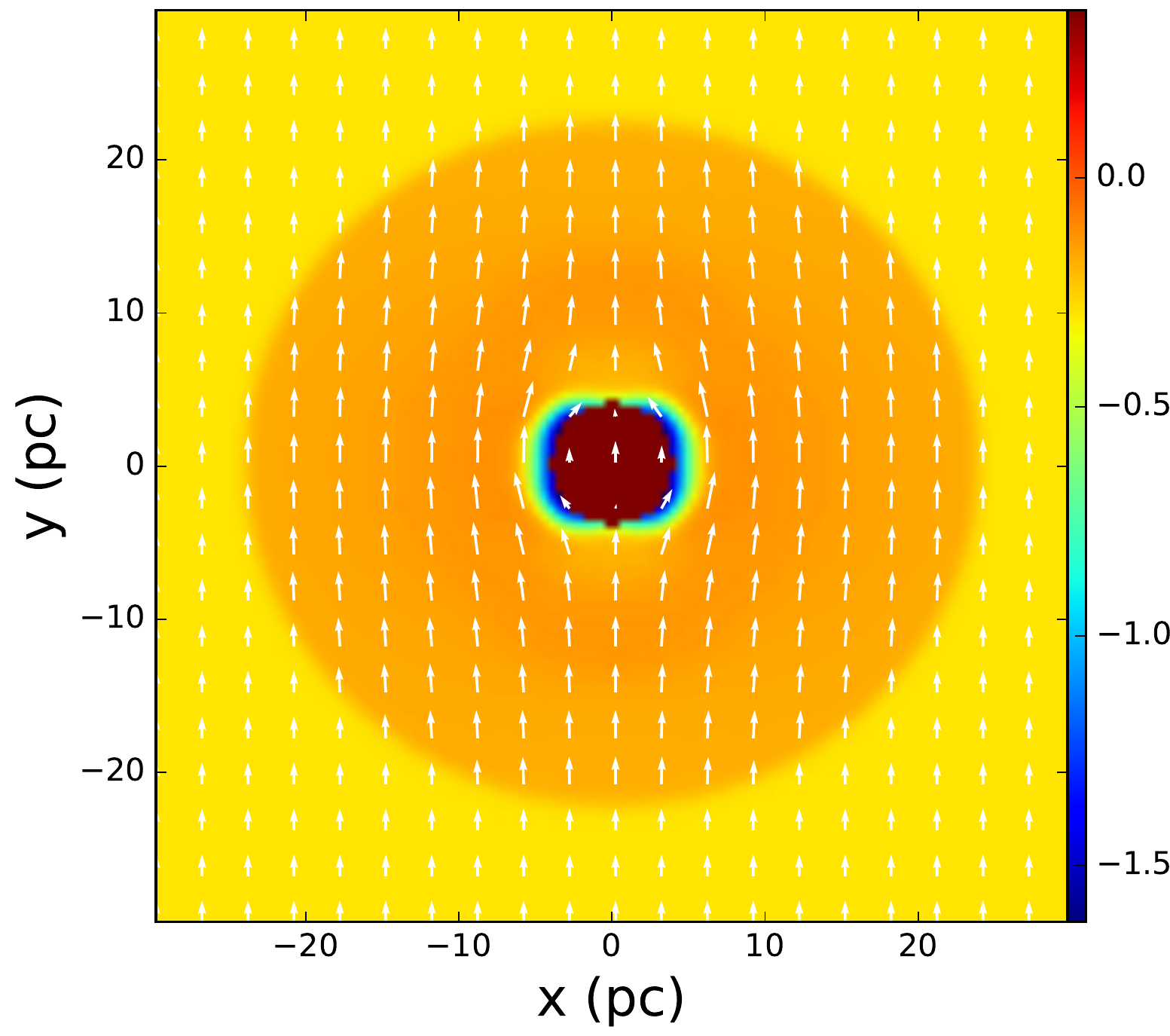}
    \includegraphics[width=0.325\textwidth]{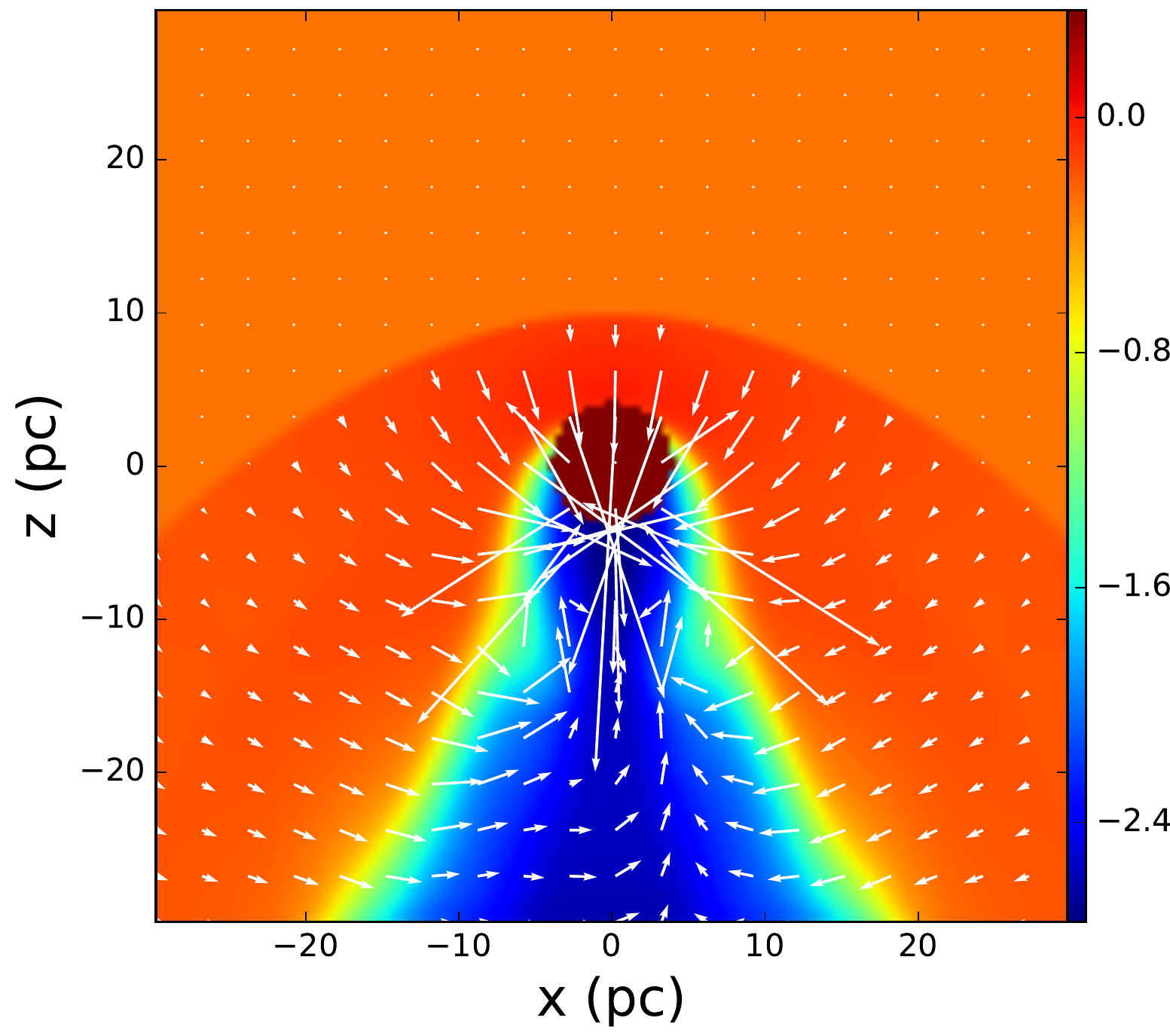}
    \includegraphics[width=0.325\textwidth]{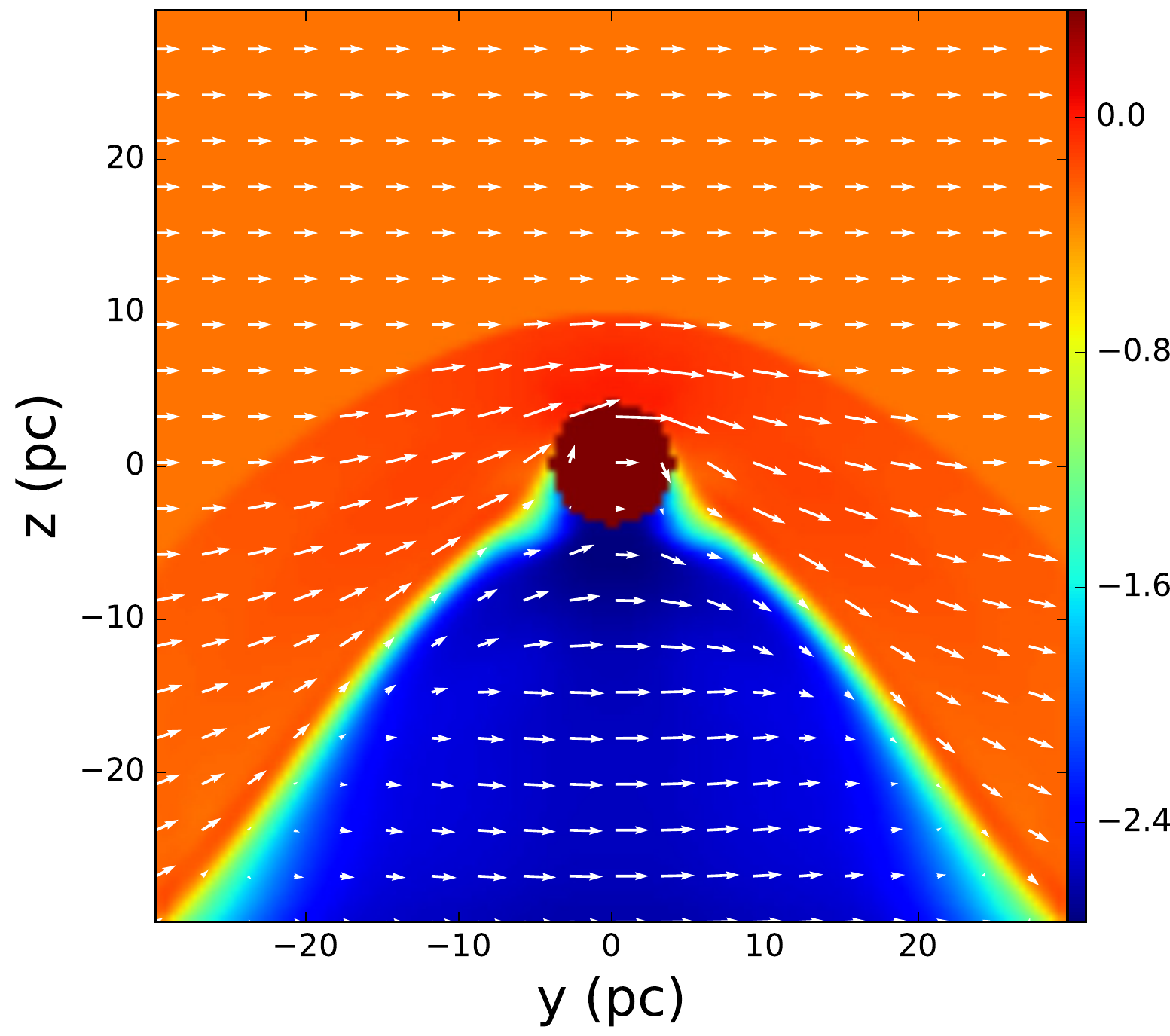}\newline
    \includegraphics[width=0.325\textwidth]{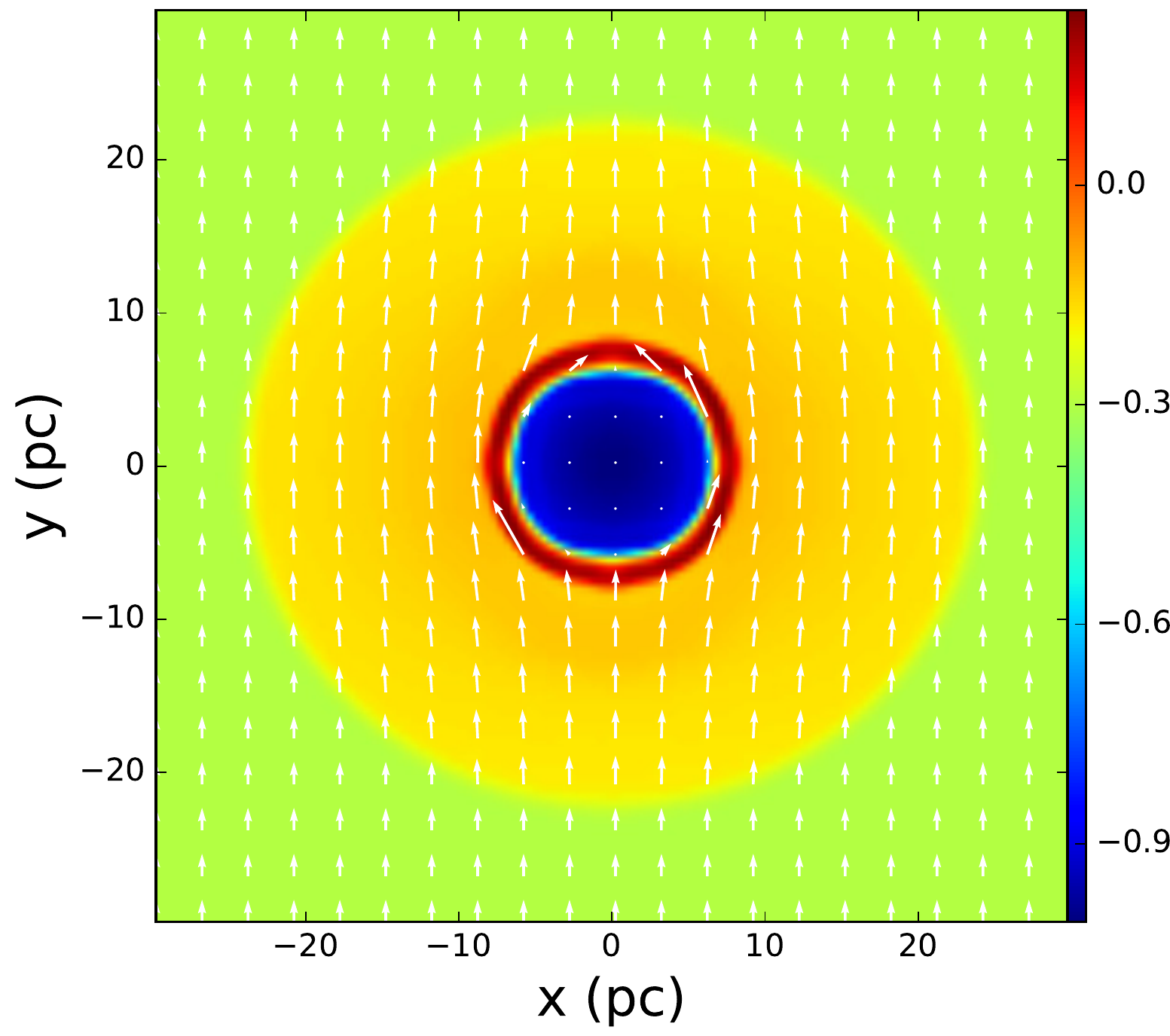}
    \includegraphics[width=0.325\textwidth]{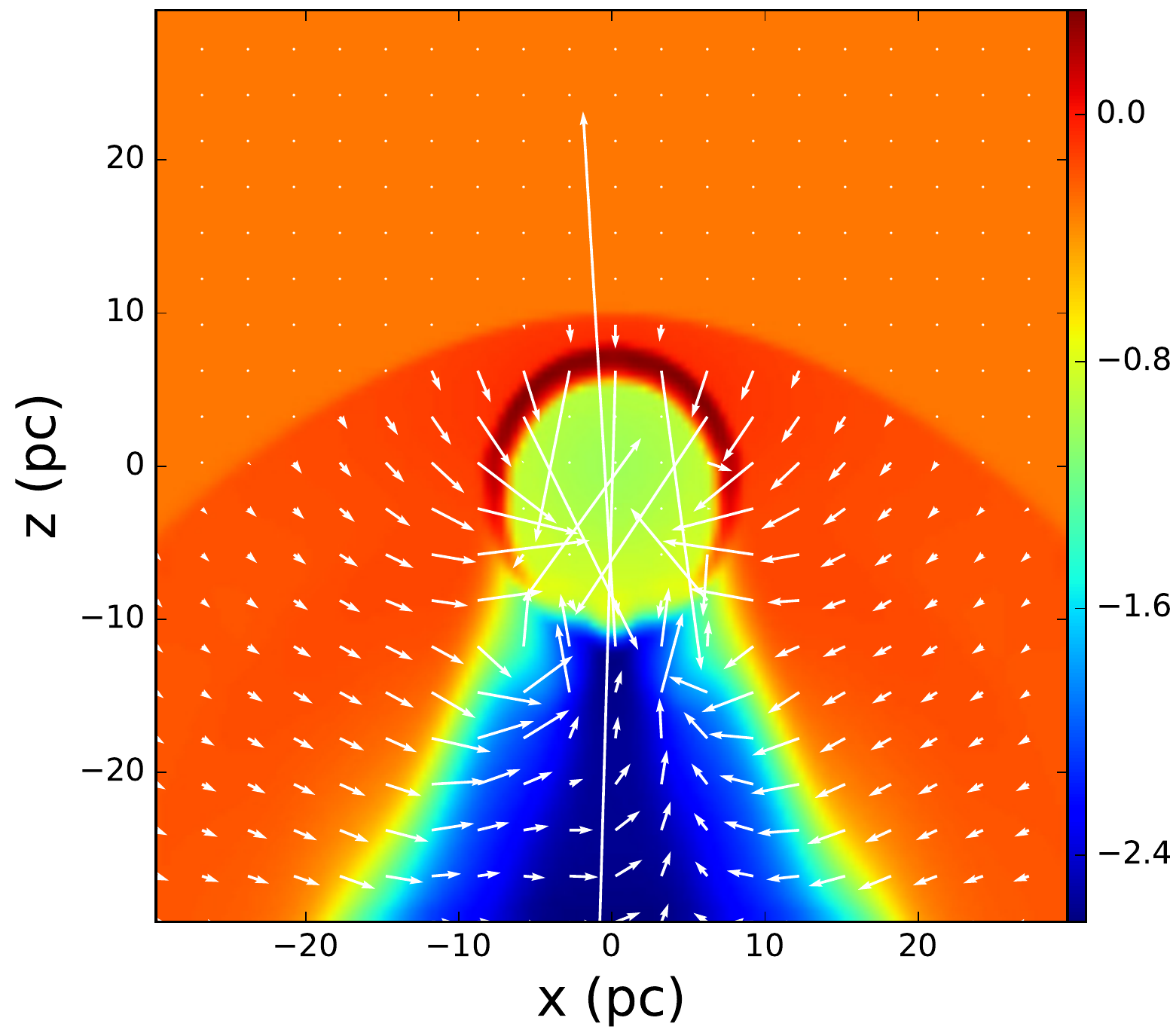}
    \includegraphics[width=0.325\textwidth]{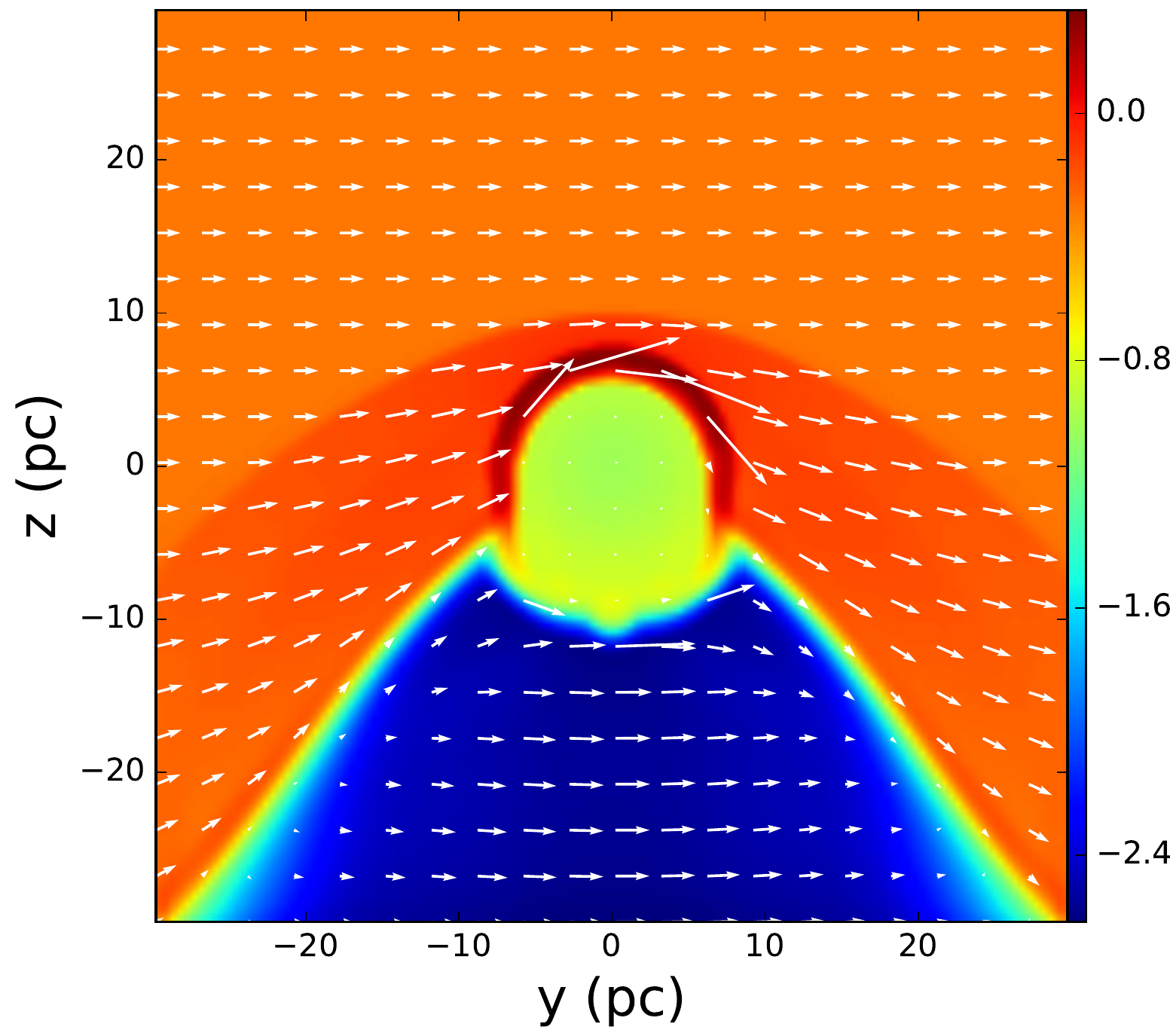}\newline
    \includegraphics[width=0.325\textwidth]{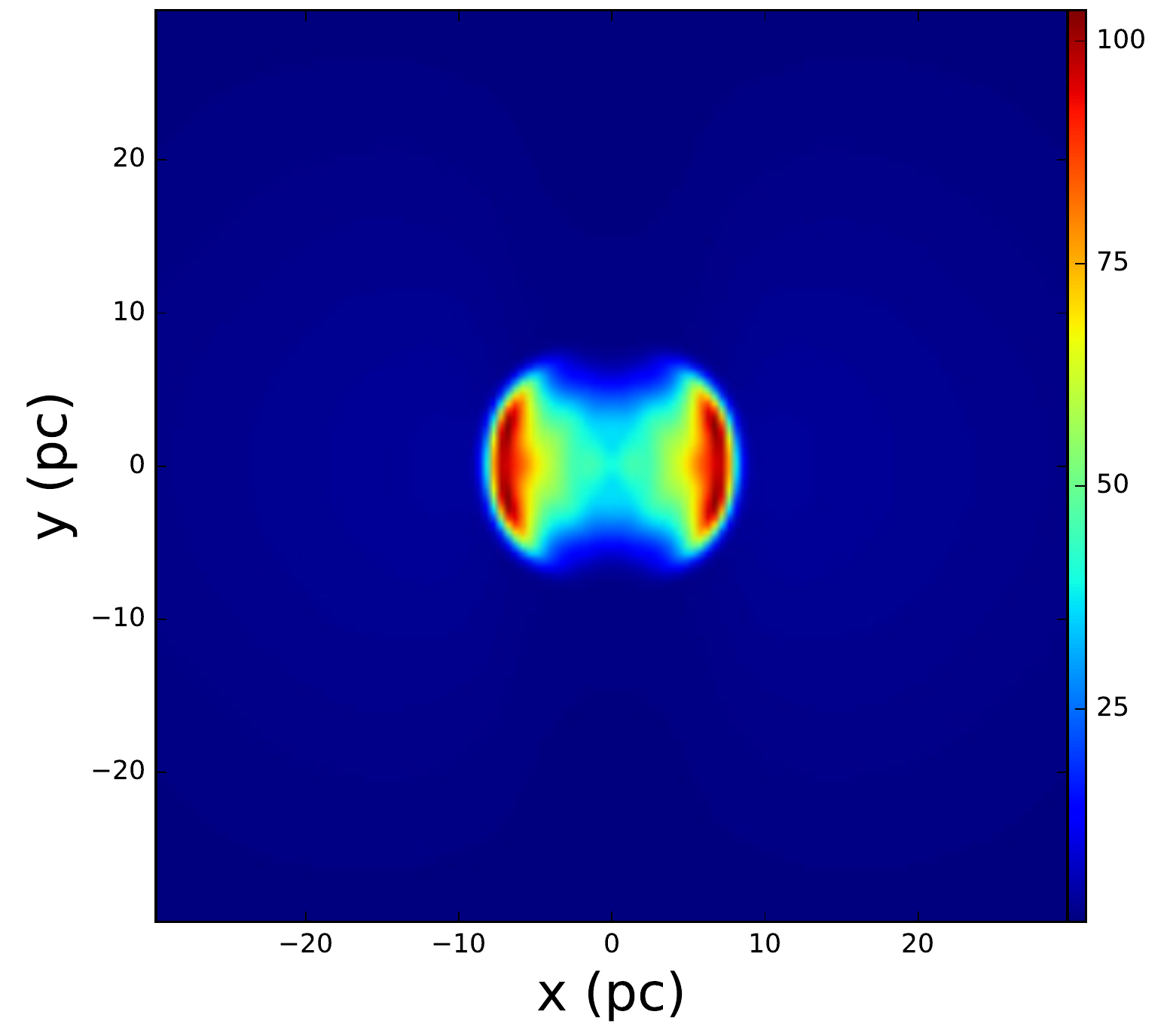}
    \includegraphics[width=0.325\textwidth]{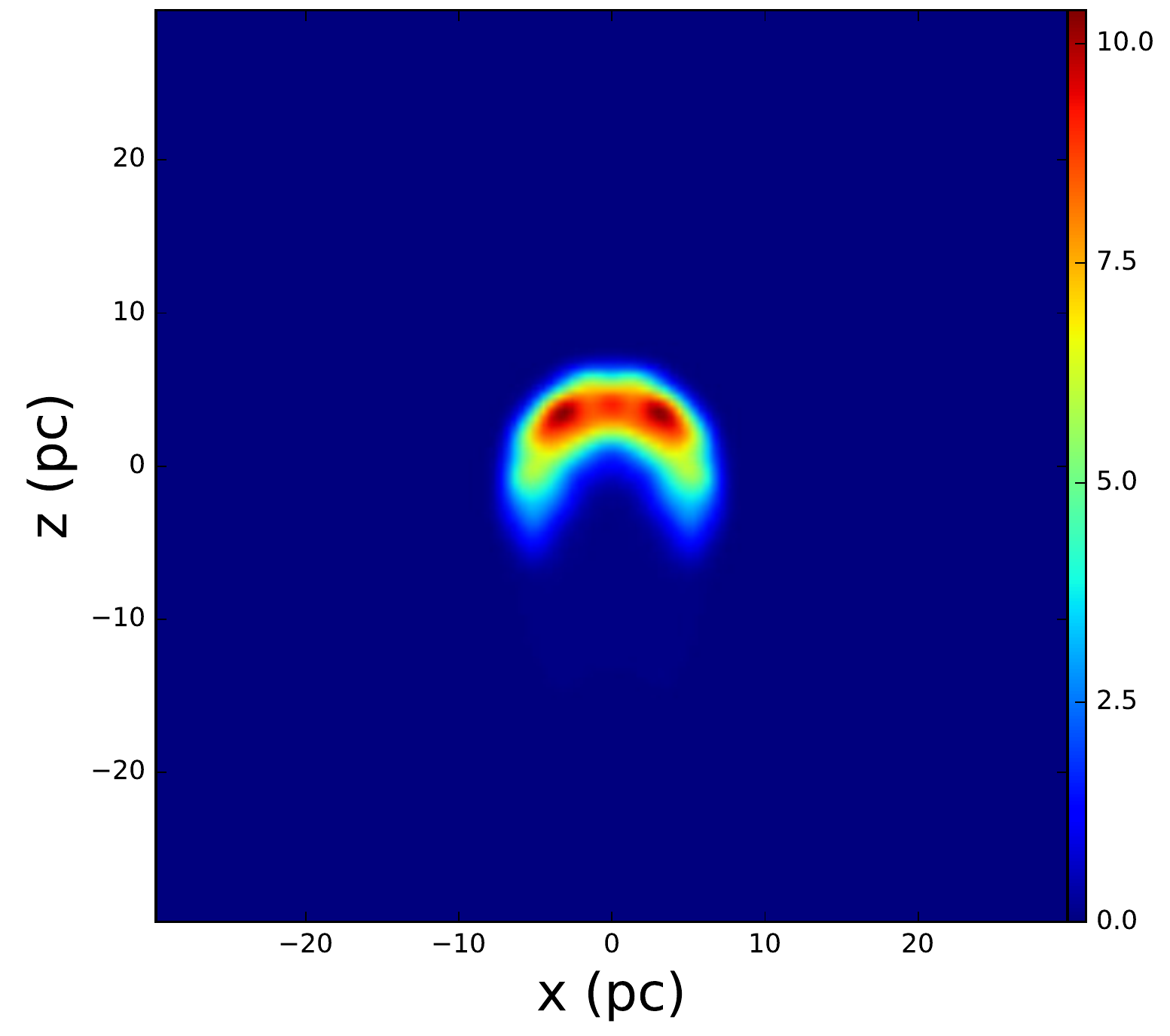}
    \includegraphics[width=0.325\textwidth]{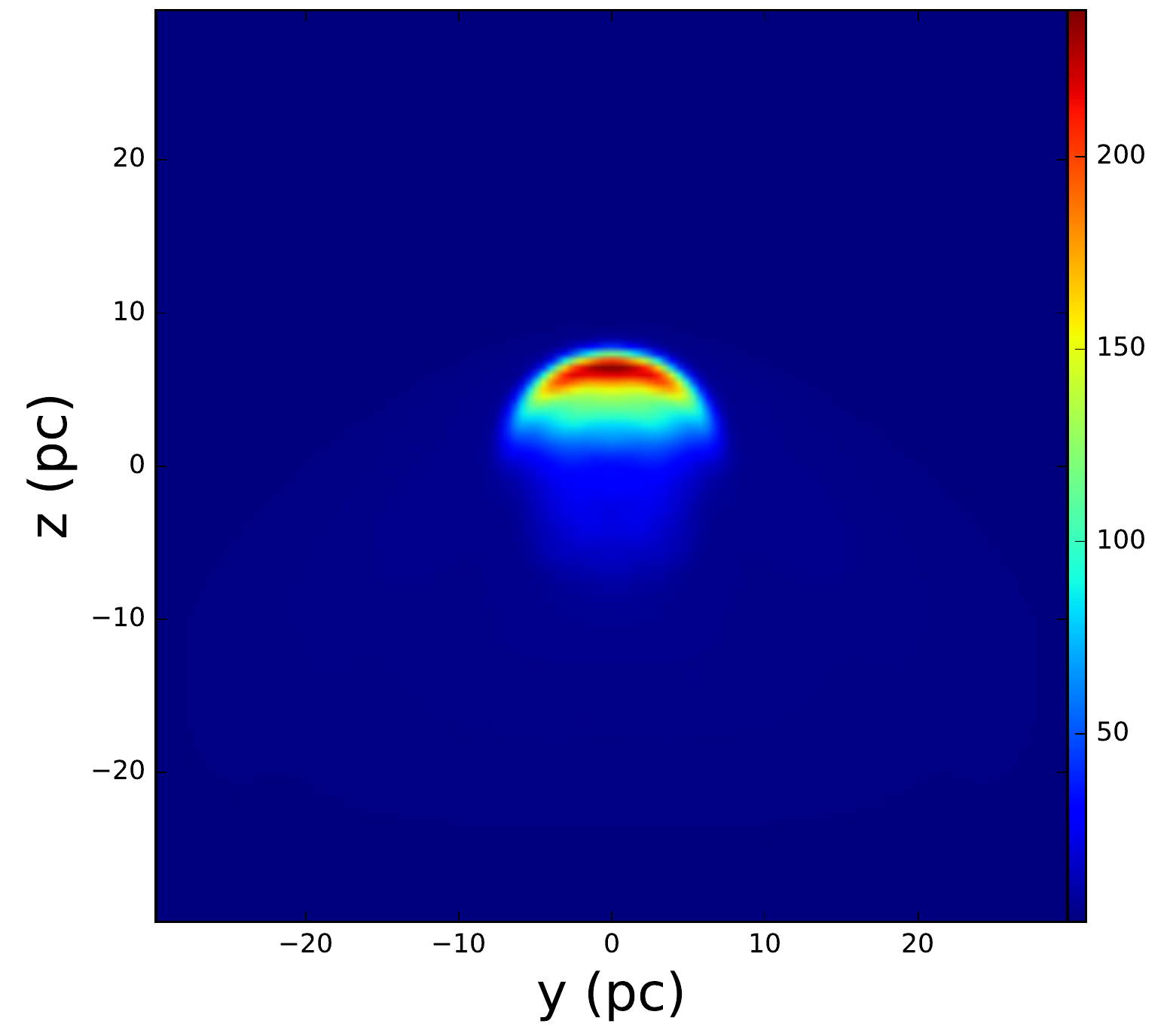}\newline
    \includegraphics[width=0.325\textwidth]{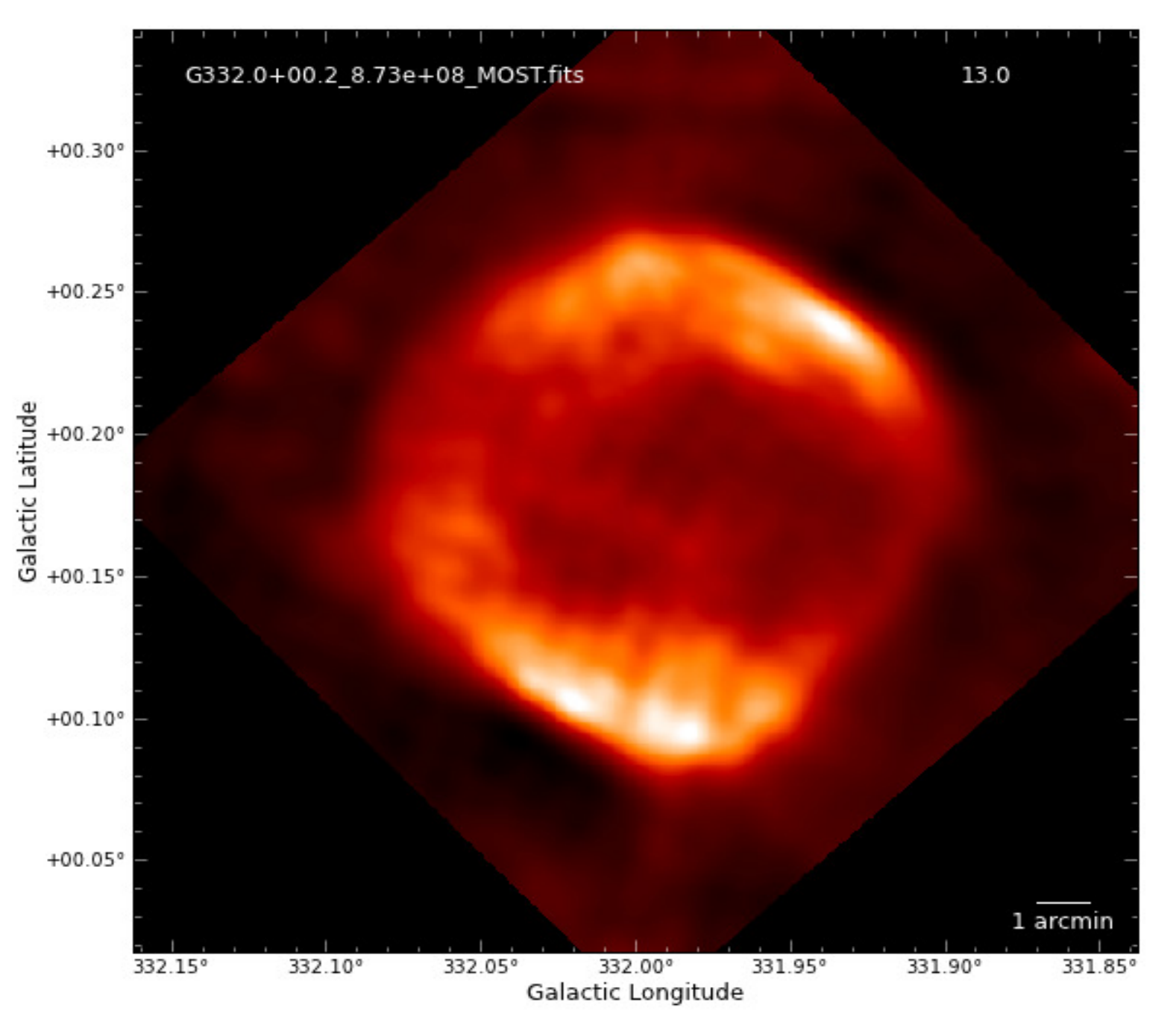}
    \includegraphics[width=0.325\textwidth]{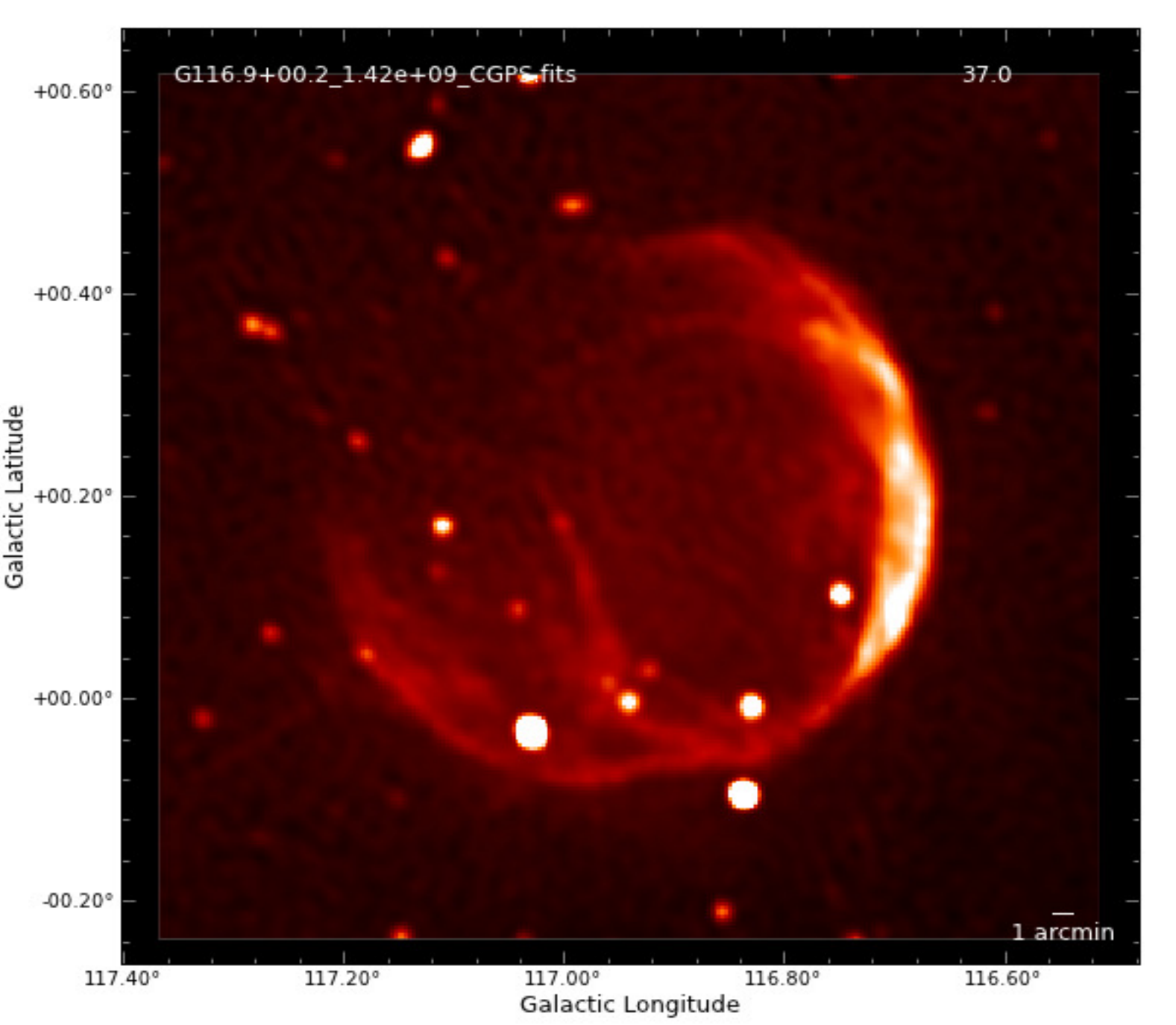}
    \includegraphics[width=0.325\textwidth]{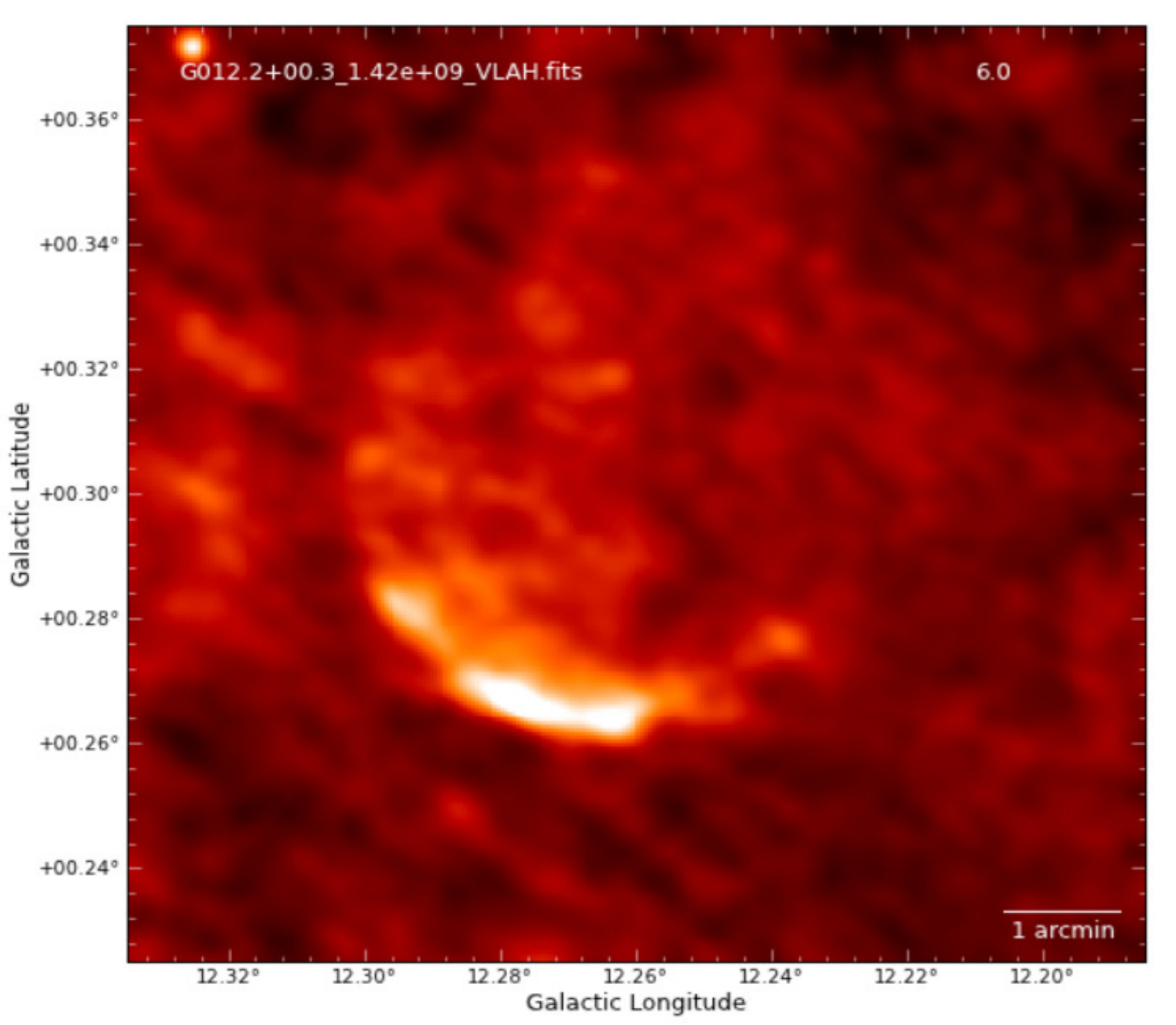}
    \caption{Simulation images assuming the velocity is perpendicular to the magnetic field. Top three images show the
    stellar wind simulation results at different views. The second row shows the SNR simulation results which apply the top
    three images as the initial conditions. The third row shows the relative radio flux density converted from the second row.
    The last row shows the real observed radio images of SNRs, G332.0+0.2, G116.9+0.2 and G12.2+0.3 \citep{West2016}.
    The three SNRs are bilateral symmetric, unilateral large-radian and unilateral small-radian, respectively. In the top two rows,
    the colorful patterns indicate the density distribution with a unit of log(cm$^{-3}$). The length and the direction of the
    white arrows respectively indicate the intensity and the direction of magnetic field.}
\label{fig:per}
\end{figure*}

The perpendicular simulation is shown in Figure~\ref{fig:per}.
The top panels show the initial conditions at three directions.
It is composed of two parts, the surrounding environment and the inner supernova explosion region.
The surrounding environment results from the stellar wind evolution and the inner's physics status is calculated based on the
work of \citet{Leahy2017a}.
The initial magnetic field and the progenitor velocity are set to follow the y-axis and z-axis respectively.
This leads to an obvious bow structure in y-z plane and the very chaotic magnetic field in x-z plane.
To make the patterns clearer, the white arrows and the pattern colors are set with different scales in different images.
The values labeled on the color bar are absolute, so they can be used to compare the densities in different images.

\begin{figure}
    \centering
    \includegraphics[width=0.4\textwidth]{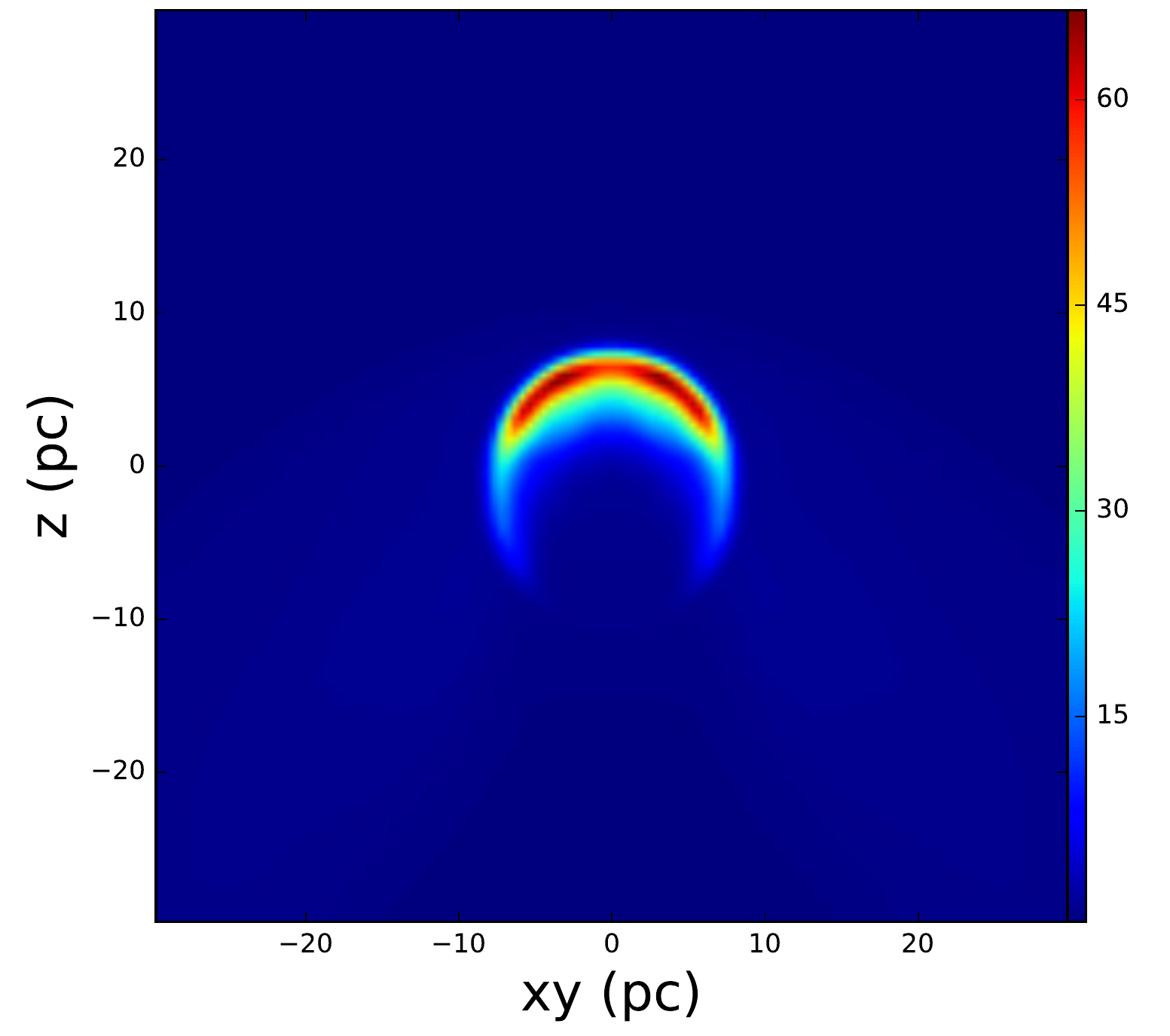}
    \includegraphics[width=0.4\textwidth]{G116-eps-converted-to.pdf}
    \caption{The simulated radio image after rotating $45\degr$ along z-axis and the observed radio image of SNR G116.9+0.2
    \citep{West2016,Tian2006}.}
\label{fig:45deg}
\end{figure}

\begin{figure*}
    \centering
    \includegraphics[width=0.9\textwidth]{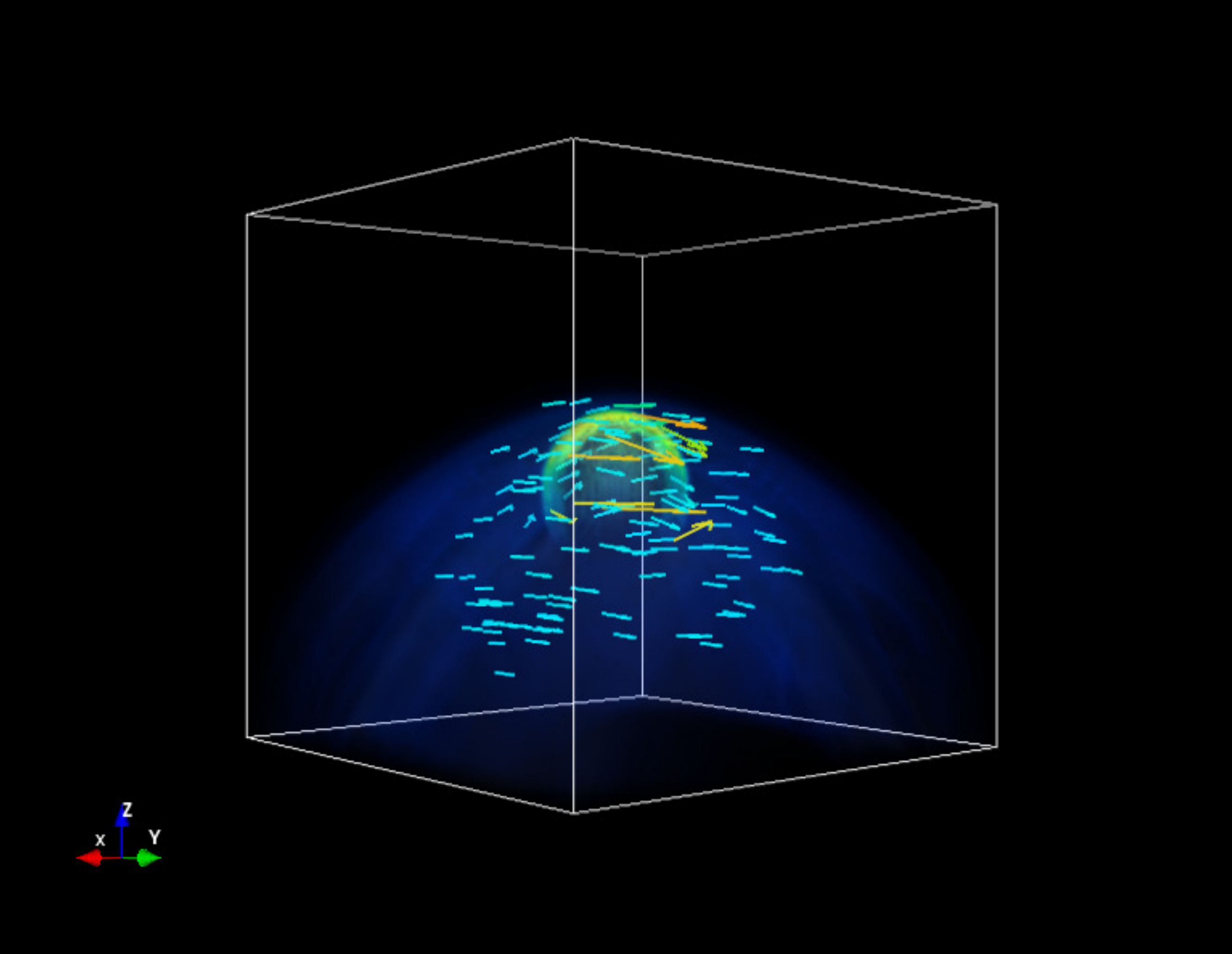}
    \caption{The simulation 3D image after rotating $50\degr$ along z-axis from the x-z plane. If we rotate it
    $45\degr$, the middle two vertical outlines will overlap with each other. Thus, we rotate a little more to make it more distinct.
    The colorful patterns indicate the relative radio flux density. The yellow shows the high flux density. The arrows show the
    magnetic field. A more yellow arrow means the larger magnetic intensity. (This figure is available online as an animation.)}
\label{fig:3D}
\end{figure*}

\begin{figure*}
    \centering
    \includegraphics[width=0.45\textwidth]{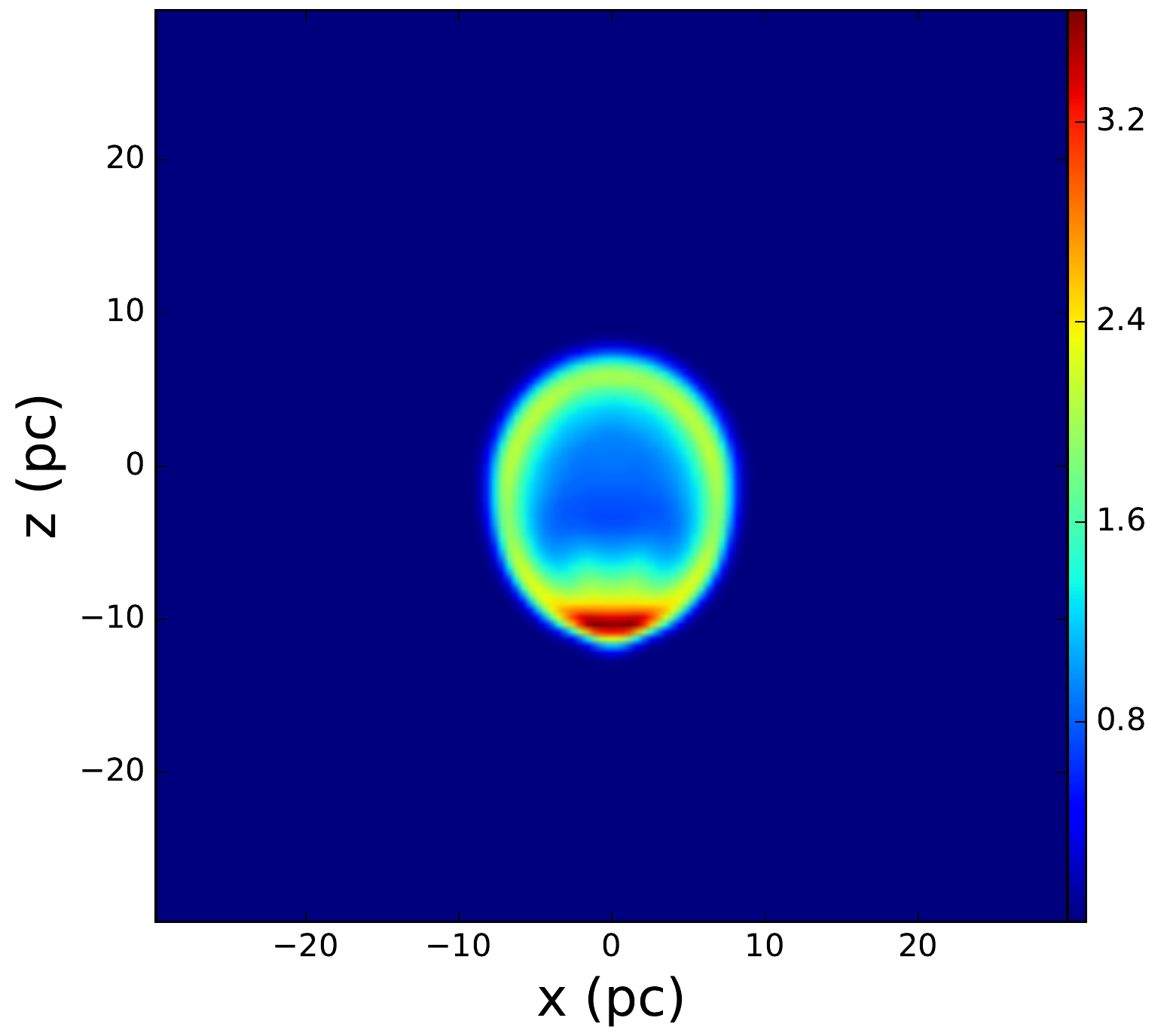}
    \includegraphics[width=0.45\textwidth]{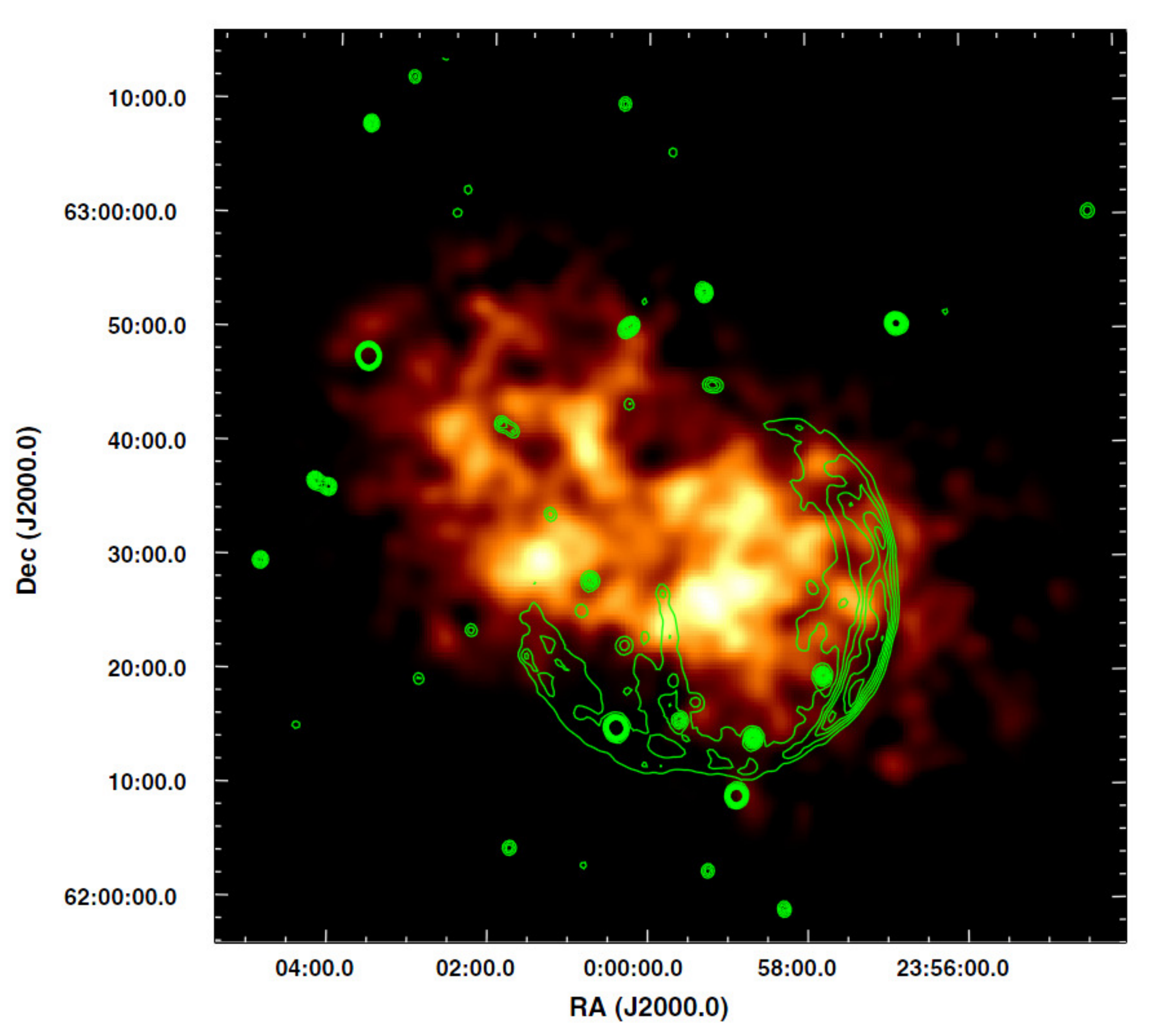}
    \caption{\textit{Left:} the relative temperature distribution in x-z plane. \textit{Right:} ASCA (Advanced Satellite for
    Cosmology and Astrophysics) X-ray image of G116.9+0.2 with CGPS (Canadian Galactic Plane Survey) radio contours overlaid (from
    \citet{Pannuti2010}).}
\label{fig:X}
\end{figure*}

The second row of Figure~\ref{fig:per} shows the SNR simulation results after 1200 years.
If we add the initial 650 years, then the age of this artificial SNR is 1850 years.
The radio morphologies, shown in the third row, are a little surprising, especially in x-z plane.
Our simulations can simultaneously result in the bilateral symmetric, the unilateral big-radian and small-radian SNRs.
As a comparison, three real SNRs \citep{West2016} are shown in the bottom panels of Figure~\ref{fig:per}.
This proves that three kind of SNRs may originate from same a progenitor, and their morphologies depend on the view
angle at which we see them.
The bilateral symmetric SNRs have been well studied by simulations and observations \citep{Gaensler1999,Petruk2009a},
but there are still many ambiguities for unilateral SNRs.
Here we show the images toward three directions, but in fact the SNR morphology varies following different view angle.
We take SNR G116.9+0.2 as an example here.
If we rotate $45\degr$ along the z-axis, we can get a unilateral bigger-radian morphology SNR in z-xy plane (see
Figure~\ref{fig:45deg}), which is more similar to the SNR G116.9+0.2.
Moreover, the magnetic field of G116.9+0.2 is parallel to the shell \citep{Sun2011} in the polarization observation,
which is totally different from the result in x-z plane.
However, if we rotate $45\degr$ along the z-axis, the magnetic field becomes similar to the observation (see Figure~\ref{fig:3D}).
The X-ray emission region of G116.9+0.2 is extended away from the radio shell \citep{Pannuti2010}, which is also revealed
by our simulation (see Figure~\ref{fig:X}).
In the left panel of Figure~\ref{fig:X}, the bottom high-temperature region is low-density comparing with the middle panel of
the second row of Figure~\ref{fig:per}, which hints it is a high-temperature low-density region full of ionized gas.
This is an appropriate environment to generate X-ray emission by bremsstrahlung mechanism.
Therefore it is possible that the high speed of the progenitor leads to the extensive X-ray emission.
\citet{Craig1997,Yar-Uyaniker2004,West2016} have ever tried to explain the X-ray morphology, but have not come to the conclusion.
A more specific simulation for SNR G116.9+0.2 will help us further understand it.

It is worthy to be mentioned that we do not add a magnetic field gradient or a density gradient at the beginning.
Even if the initial ISM is uniform, we can still obtain various morphologies.
In other words, the radio morphology is not only dependent on the initial ISM distribution.
Therefore, it is unreasonable to estimate the initial magnetic field or density distribution before the progenitor formation
based on the radio morphology of a SNR.
Also the radio morphology should not be used to infer the large-scale magnetic field or density distribution in Milky Way,
since the local environment has been changed by the stellar wind, which leads to the difference between local and large-scale
environment.
In fact, \citet{Orlando2007} obtained similar radio morphologies based on inhomogeneous initial ISM settings,
but they did not explain the origin of such initial conditions.
\citet{vanMarle2010} took the stellar wind into consideration and explained the its influences based on HD simulations,
but did not get radio images.
Moreover, they both did not consider the motion of the progenitor.
It is well-known that most of stars are moving against the surrounding environment, so our work is a meaningful supplement to the
previous research.
In fact, aiming at particular SNRs, \citet{Vigh2011} tried to study the asymmetries of Tycho SNR,
while \citet{Schneiter2006} generated the morphology of SNR 3C 400.2 and discussed the effect of the thermal conduction.
\citet{Toledo-Roy2014} took the motion of the progenitor into consideration and well explained the
morphology of Kepler SNR by including the stellar wind.
Further, \citet{Toledo-Roy2014a} combined the X-ray and radio emissions and studied SNR G352.7−0.1 based on a MHD simulation, but they
did not consider the stellar wind and the motion of the progenitor in their study.

Figure~\ref{fig:per} has shown that the relative flux densities in different planes are different.
The flux density is low in x-z plane, and higher in x-y plane, then the highest in y-z plane.
So it is reasonable that the unilateral small-radian SNRs appear more frequent than the
unilateral big-radian SNRs, because bright SNRs are easier to be detected.
Such a derivation is supported by the statistics of the SNR morphologies (see Table.~\ref{table:stat}).
Therefore there should exist more undiscovered unilateral big-radian SNRs in our Galaxy.
The third row of Figure~\ref{fig:per} shows that the top flux density of y-z plane is about 20 times larger than that in the x-z plane,
so it is possible to detect more unilateral big-radian SNRs once we get the sensitivity 20 times better.
The fact that the number of the observed SNRs (about 300, see \citet{2014BASI...42...47G}) is much less than the theory prediction of
above 1000 by now \citep{Frail1994a,Tammann1994}, can be partly explained by the simulation results.

\begin{figure*}
    \centering
    \includegraphics[width=0.325\textwidth]{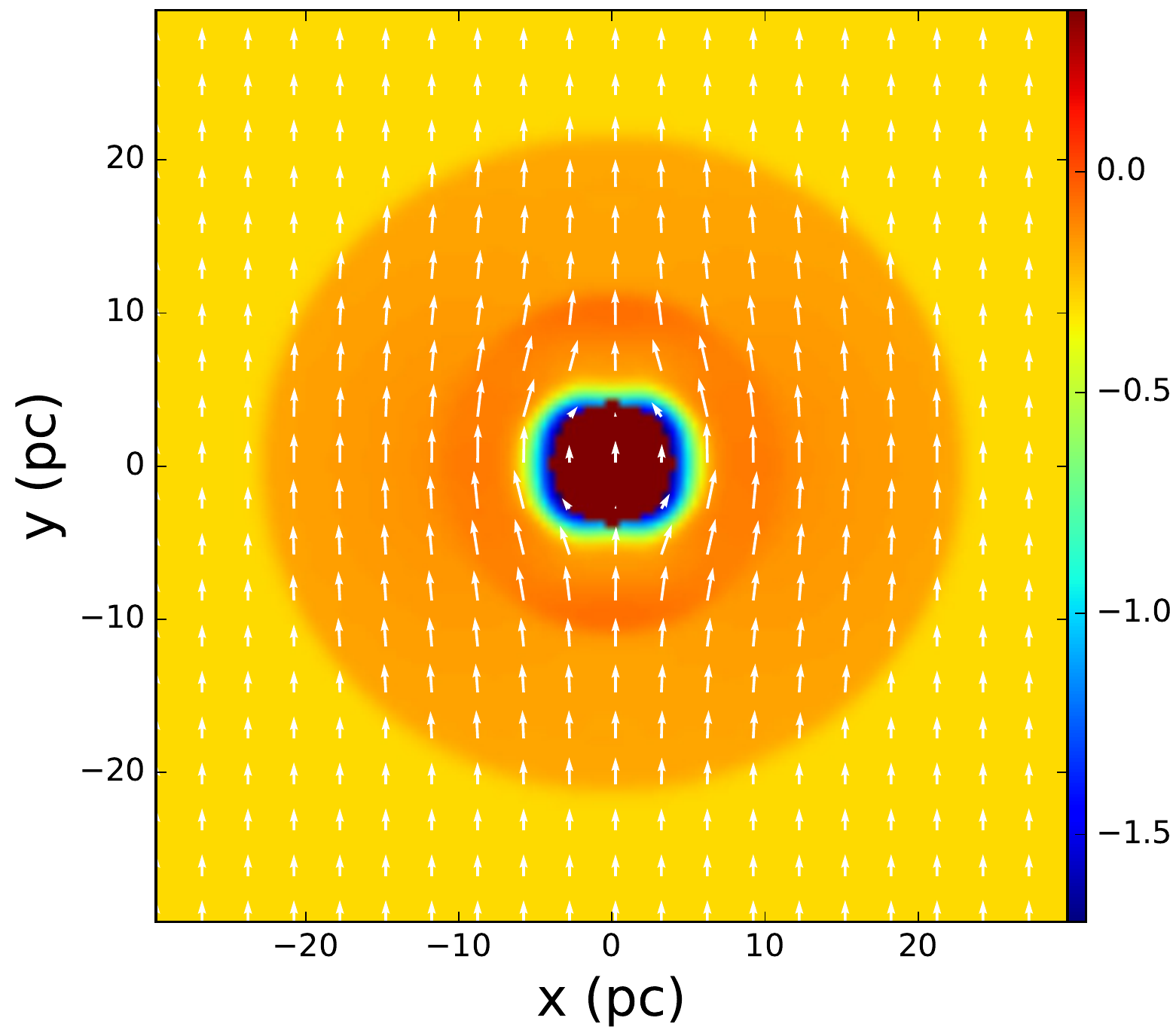}
    \includegraphics[width=0.325\textwidth]{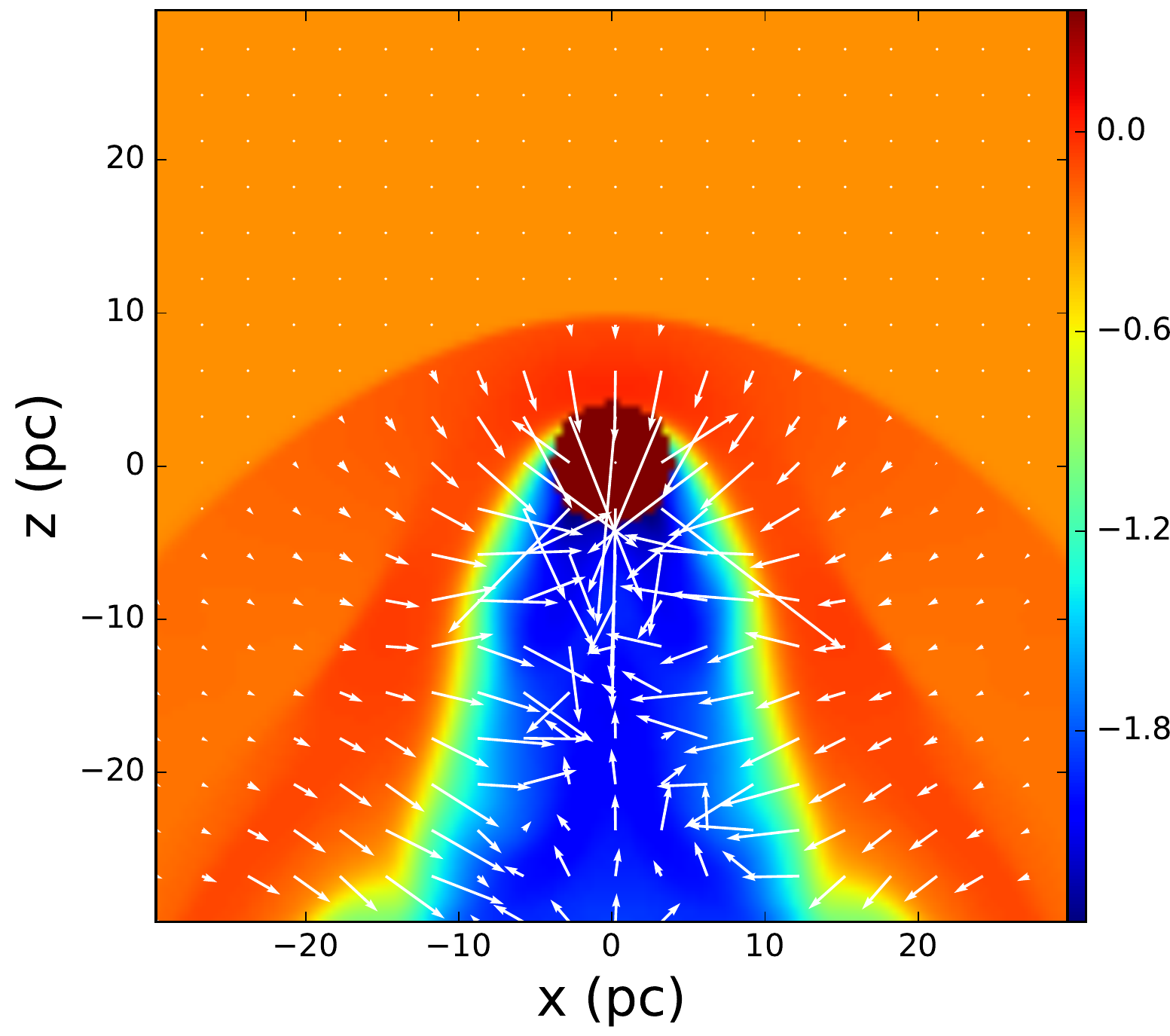}
    \includegraphics[width=0.325\textwidth]{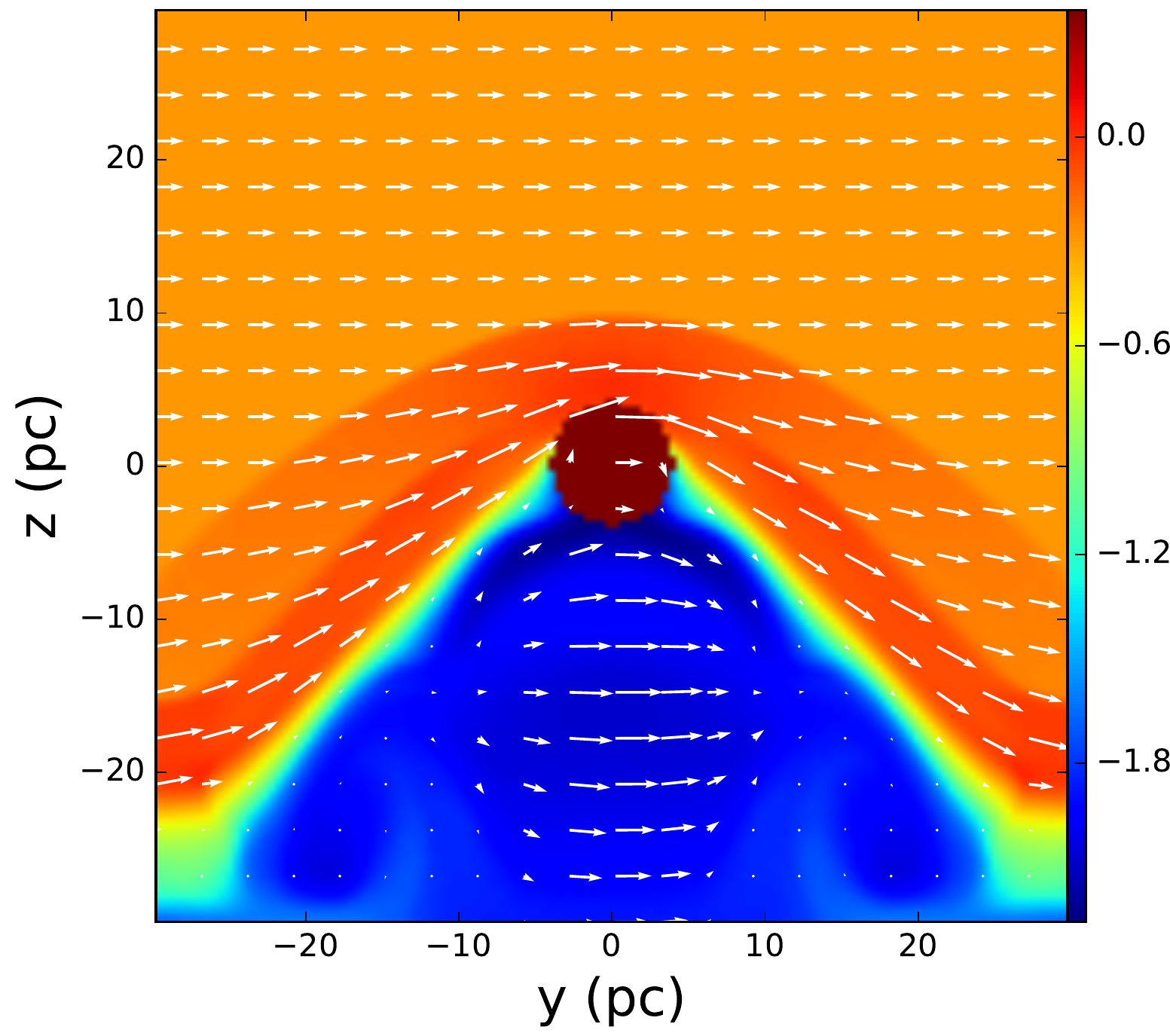}\newline
    \includegraphics[width=0.325\textwidth]{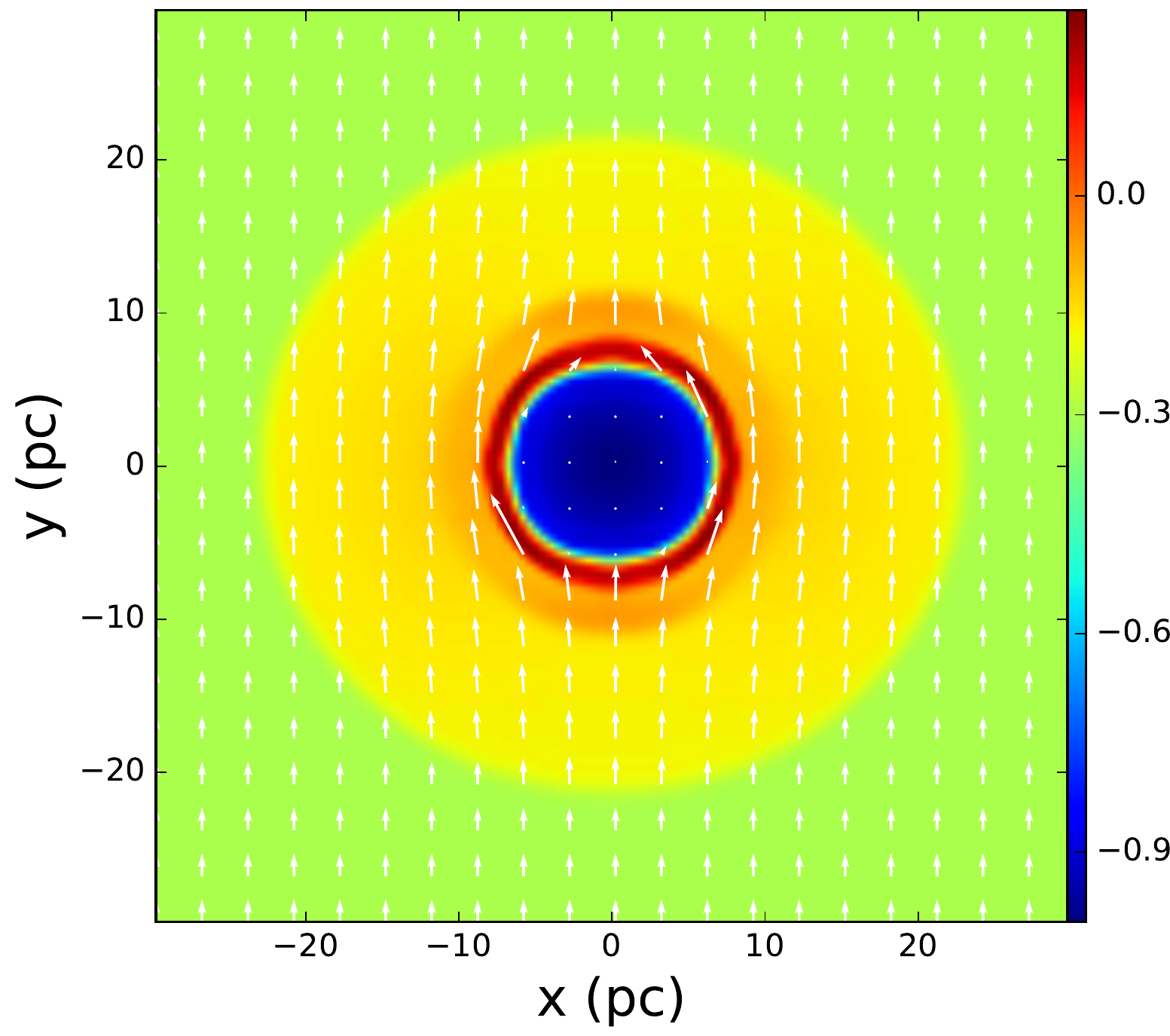}
    \includegraphics[width=0.325\textwidth]{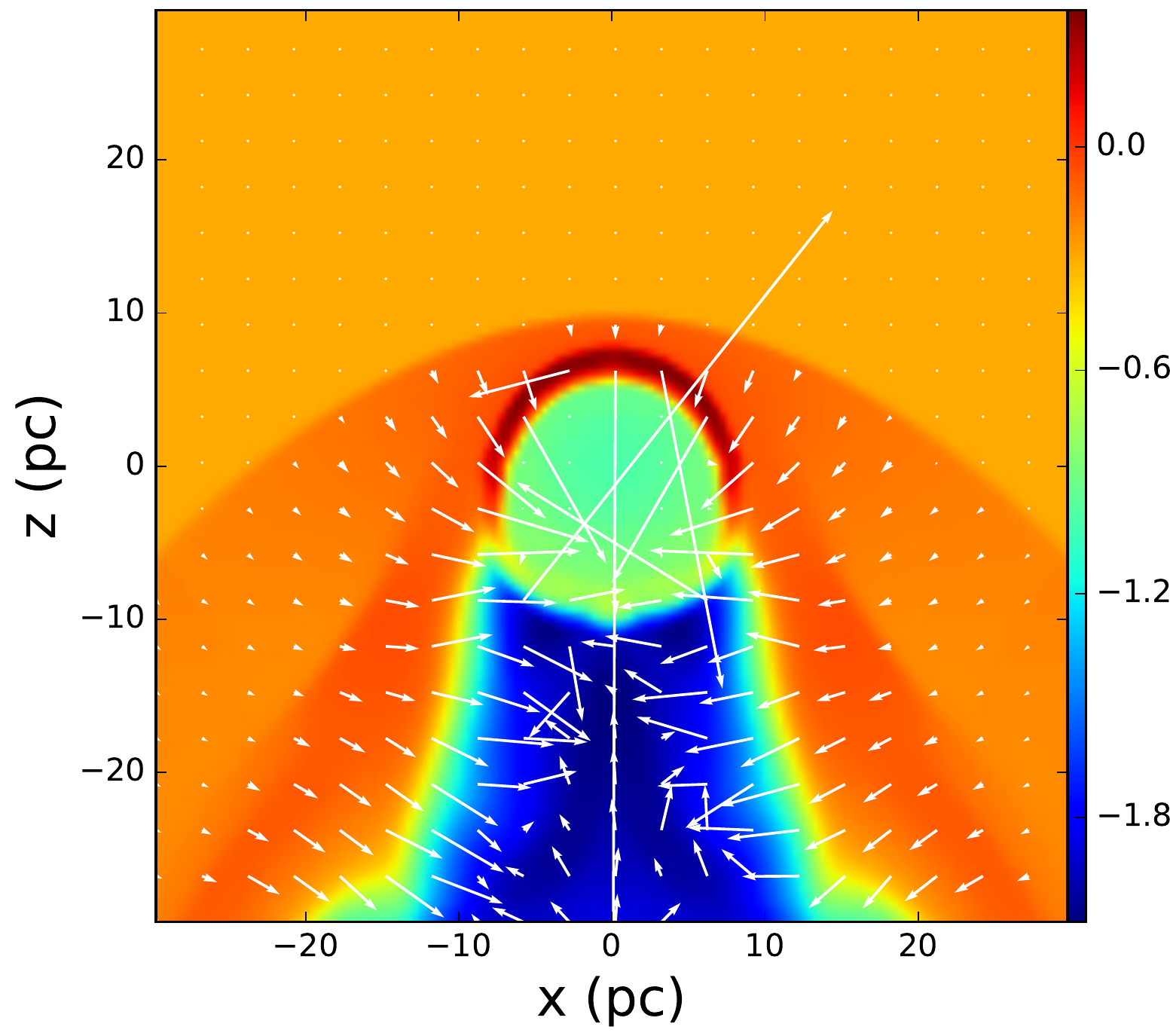}
    \includegraphics[width=0.325\textwidth]{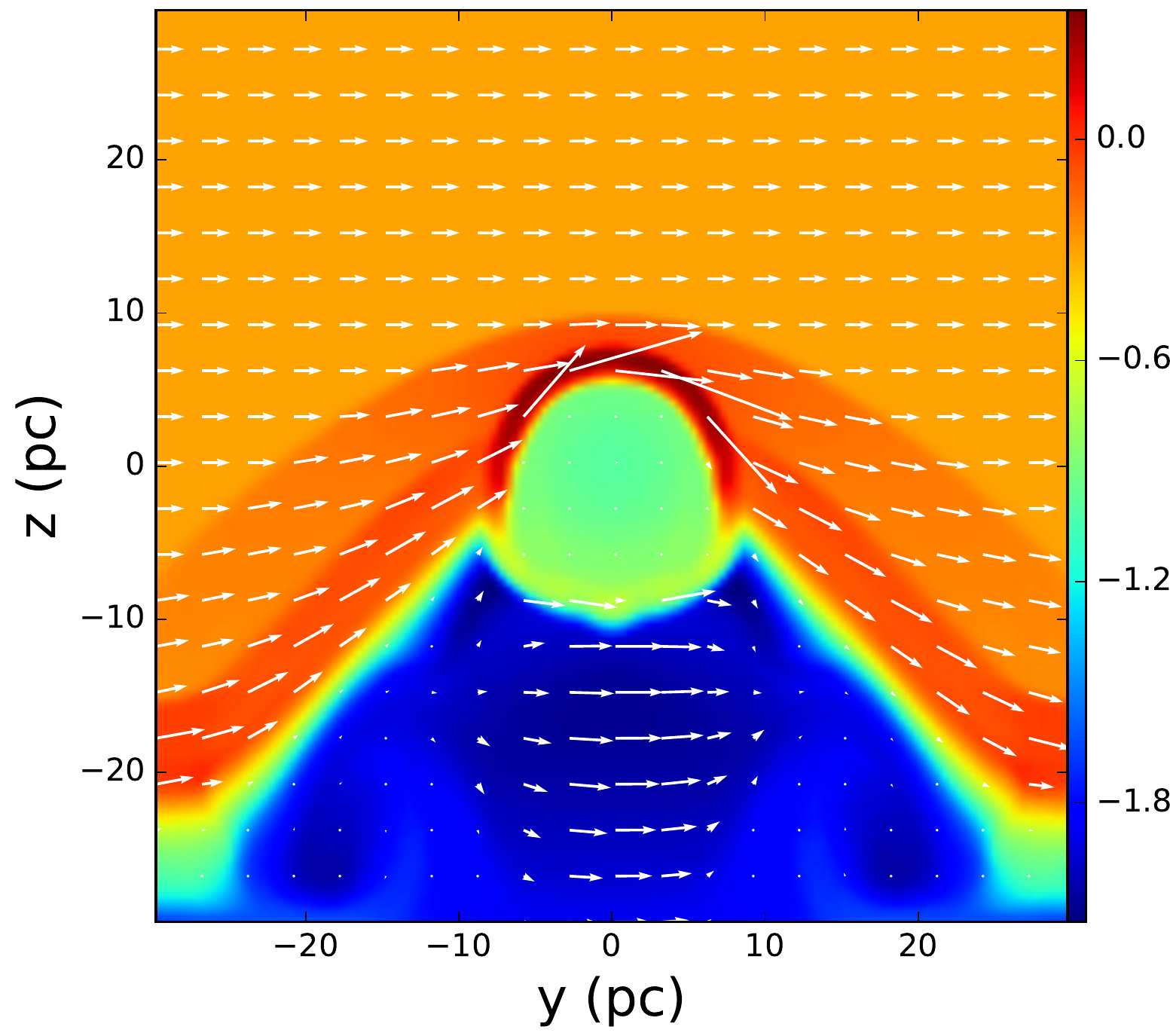}\newline
    \caption{Simulation images with thermal conductions. They are similar to the top two rows of Figure~\ref{fig:per}.
    The only difference is that the thermal conduction is included in the simulation.}
\label{fig:conduction}
\end{figure*}

We also try to check the influence of thermal conduction in the simulation,
because the thermal conduction plays an important role in the evolution of stellar wind \citep{Meyer2014}.
We apply the explicit scheme and the standard thermal conduction coefficients in the code PLUTO.
Figure~\ref{fig:conduction} shows our simulation results.
The simulation reveals that the bow shell has two layers and the magnetic field is also different from
that without the thermal conduction (see Figure~\ref{fig:per}).
\citet{Meyer2015} showed the effects on the mixing of material, which is not obvious in our work,
because we use different parameters.
The simulation including thermal conduction does not show obvious change in the density and magnetic field evolution
around the SNR.
The radio morphologies are similar to those in Figure~\ref{fig:per}, so we do not show them.
In conclusion, the thermal conduction plays a small role in the radio evolution of a SNR.

\subsection{Parallel Simulation}

\begin{figure*}
    \centering
    \includegraphics[width=0.325\textwidth]{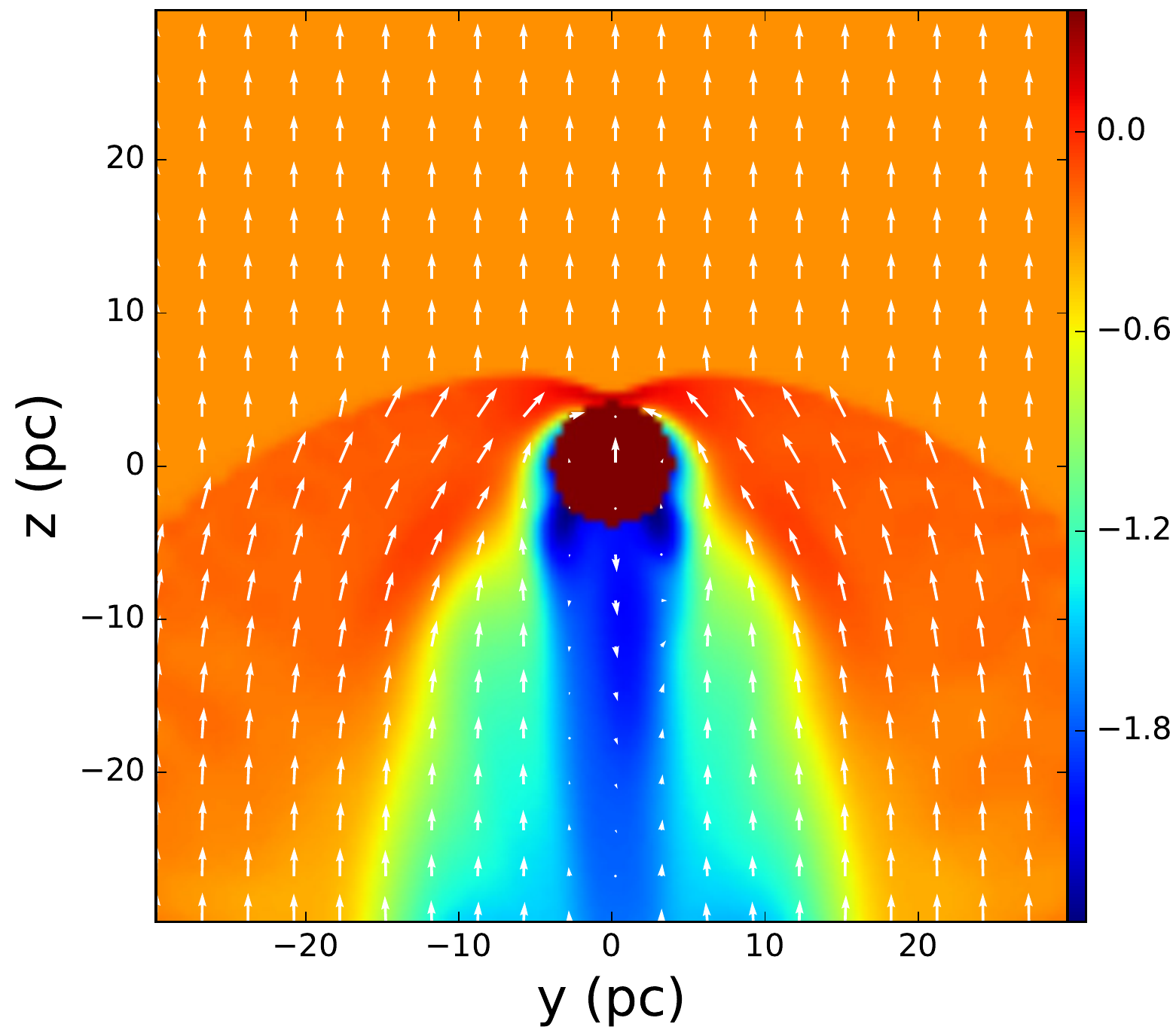}
    \includegraphics[width=0.325\textwidth]{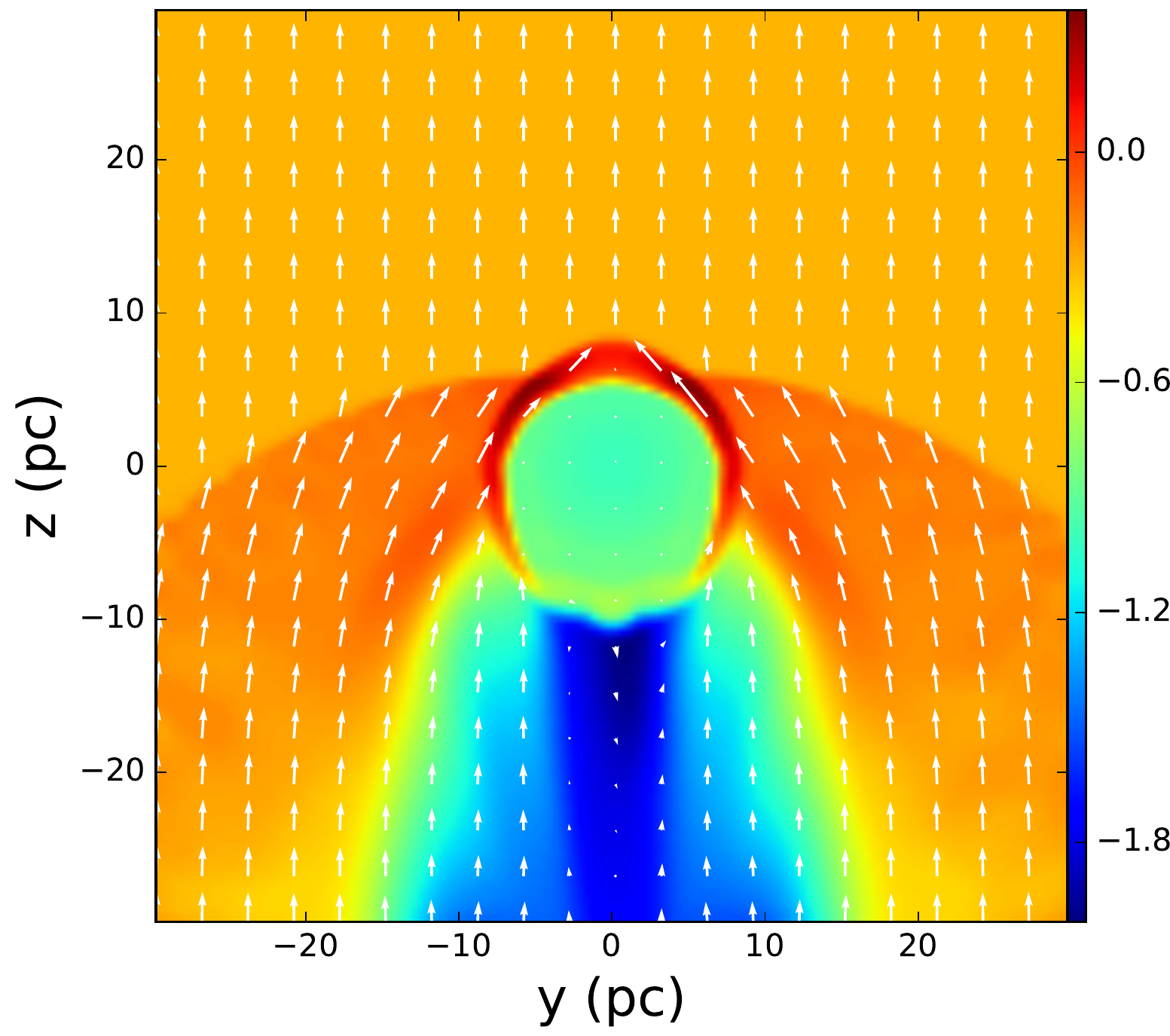}
    \includegraphics[width=0.325\textwidth]{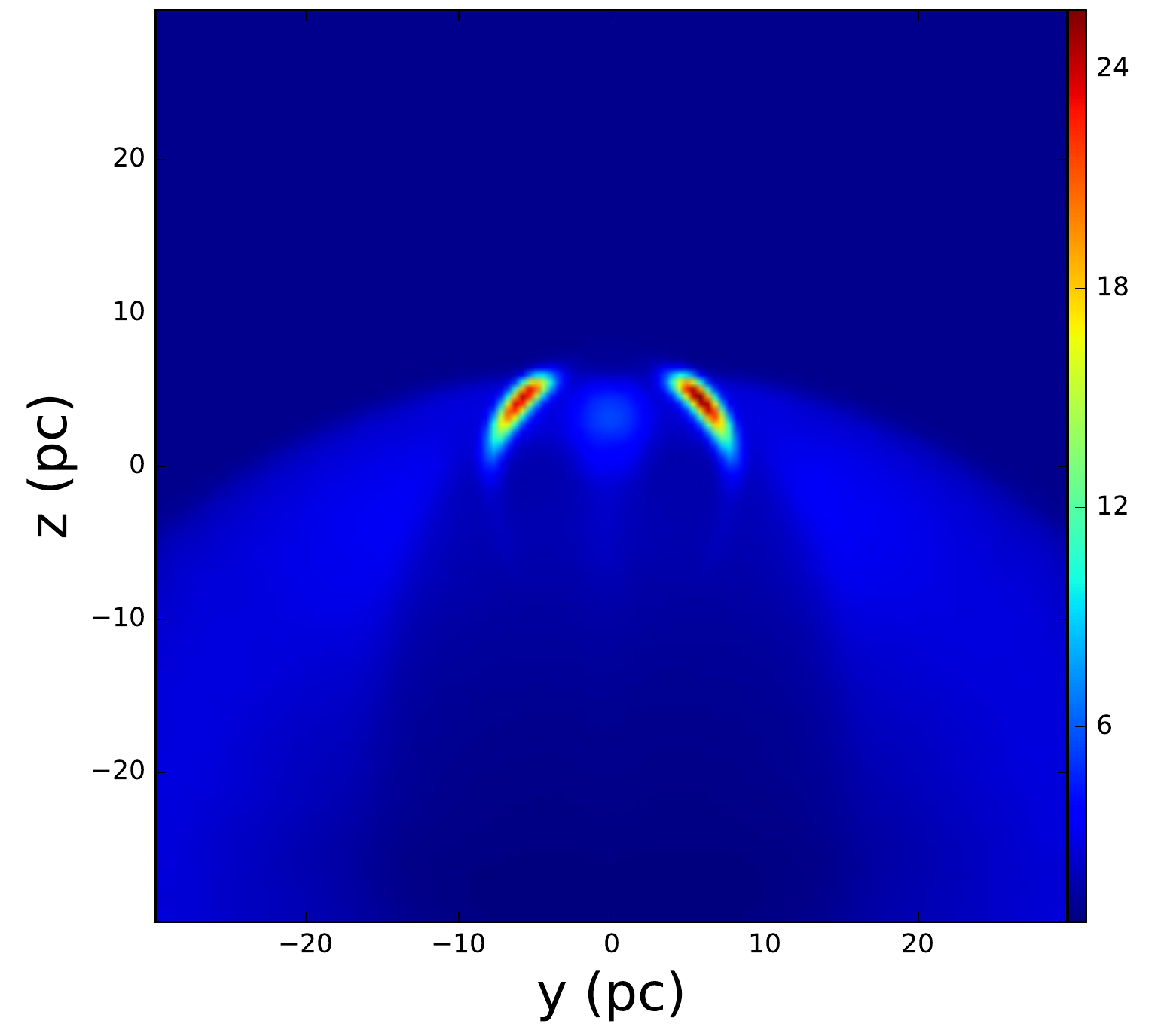}
    \caption{Simulation images assuming the velocity is parallel to the magnetic field. The left panel shows the
    stellar wind simulation result at y-z plane. The middle panel shows the SNR simulation result at y-z plane.
    The right panel shows the relative radio flux density converted from the middle panel. }
\label{fig:par}
\end{figure*}

The parallel simulation is shown in Figure~\ref{fig:par}.
All initial parameters are same as the perpendicular simulation and the age is also 1850 years.
We warn that the stellar wind region shows obvious radio emission, which is wrong, because there is no
relativistic electron in the stellar wind region and synchrotron mechanism is here not important.
However, it is impossible for us to exclude it from the radio images, because we do not know the boundary of the
relativistic electrons region.
This flaw also influences other simulation radio images.
We only show the y-z plane in Figure~\ref{fig:par}, because the x-z plane is same as the y-z plane.
Moreover, we should see a circular SNR in the x-y plane but in fact a square SNR in our simulation,
because the resolution is not high and every pixel is square.
The stellar wind simulation is time-consuming, so we selectively set a reasonable resolution.

\begin{figure*}
    \centering
    \includegraphics[width=0.325\textwidth]{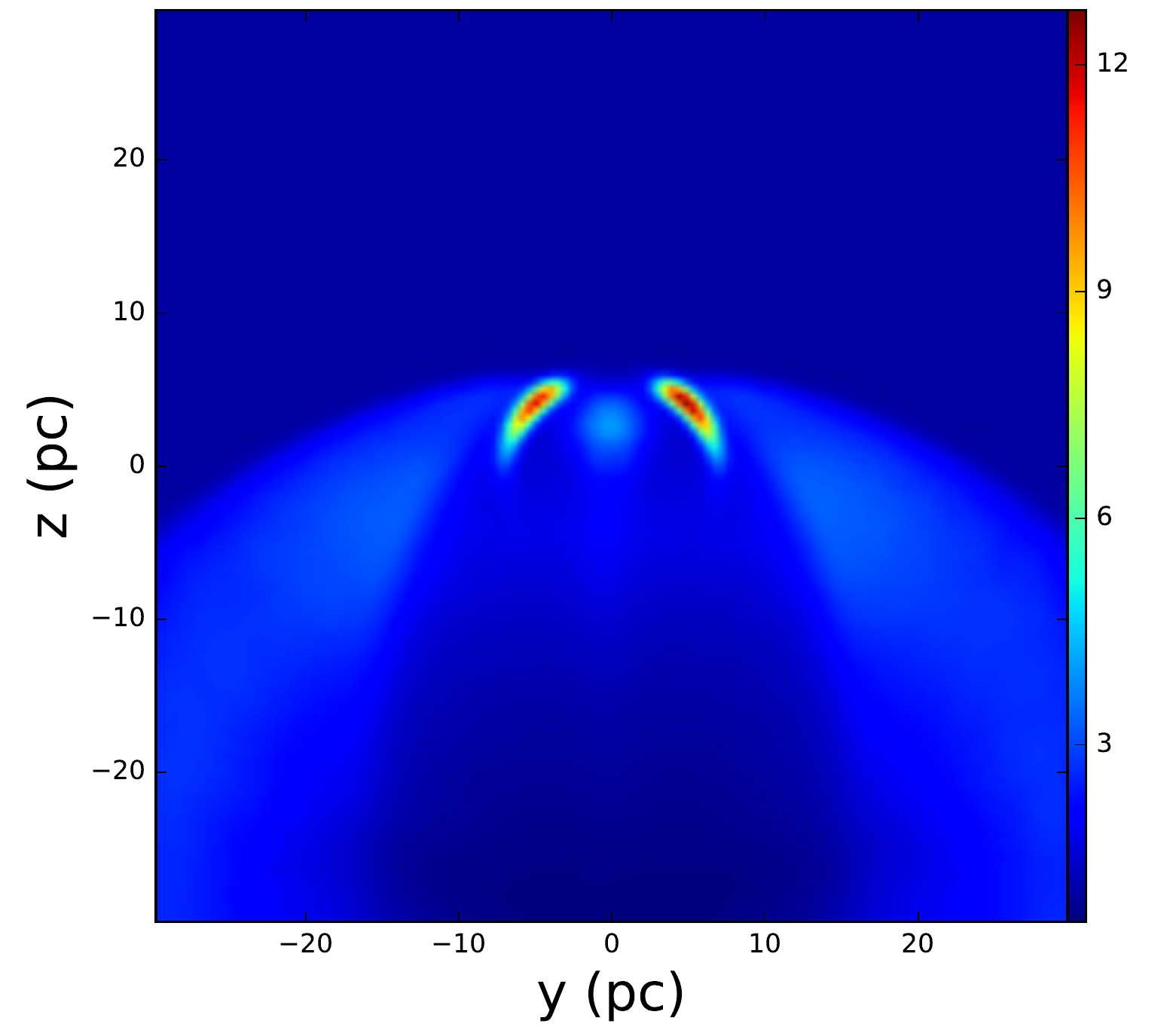}
    \includegraphics[width=0.325\textwidth]{t6_yz-eps-converted-to.pdf}
    \includegraphics[width=0.325\textwidth]{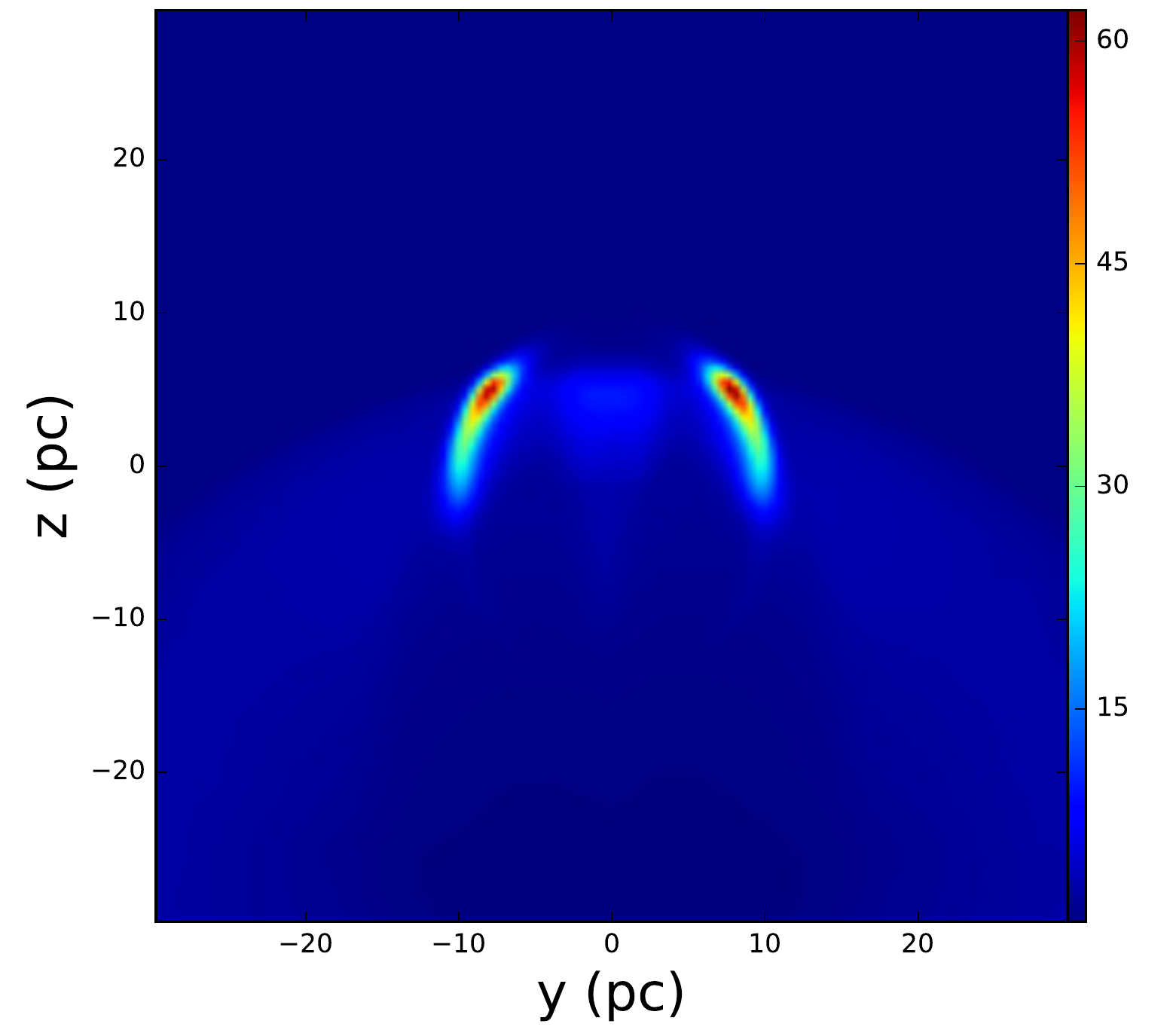}\newline
    \includegraphics[width=0.325\textwidth]{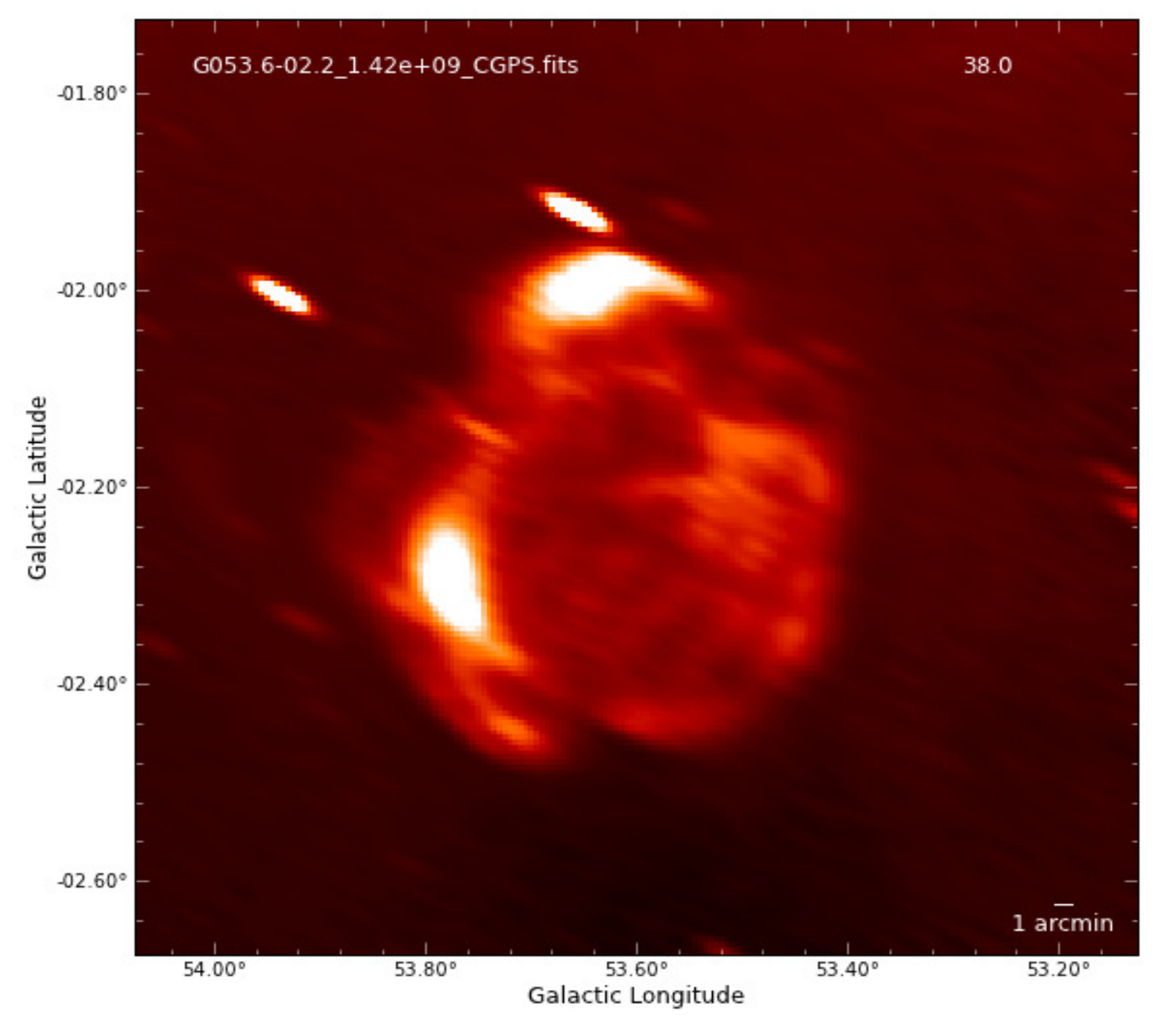}
    \includegraphics[width=0.325\textwidth]{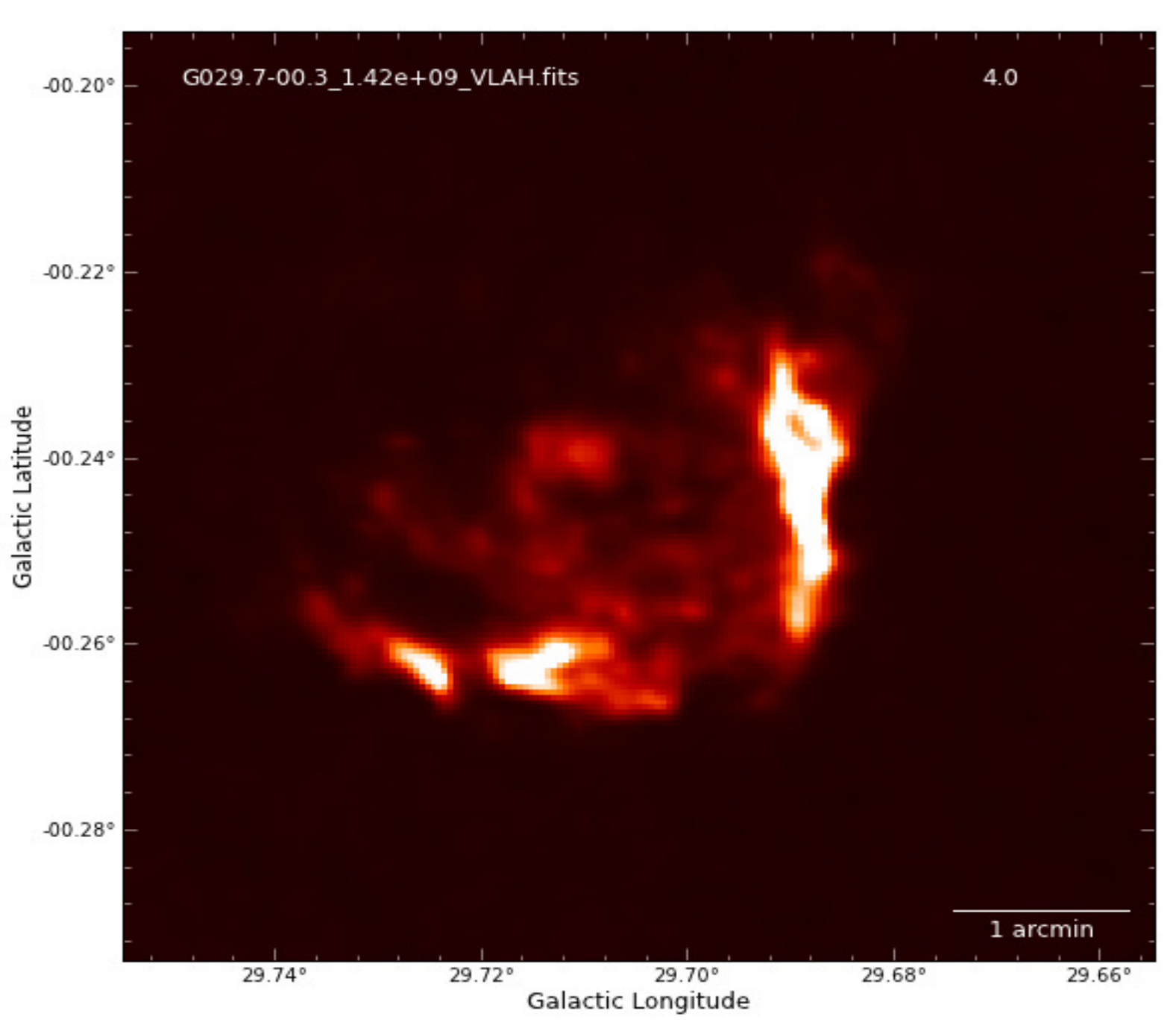}
    \includegraphics[width=0.325\textwidth]{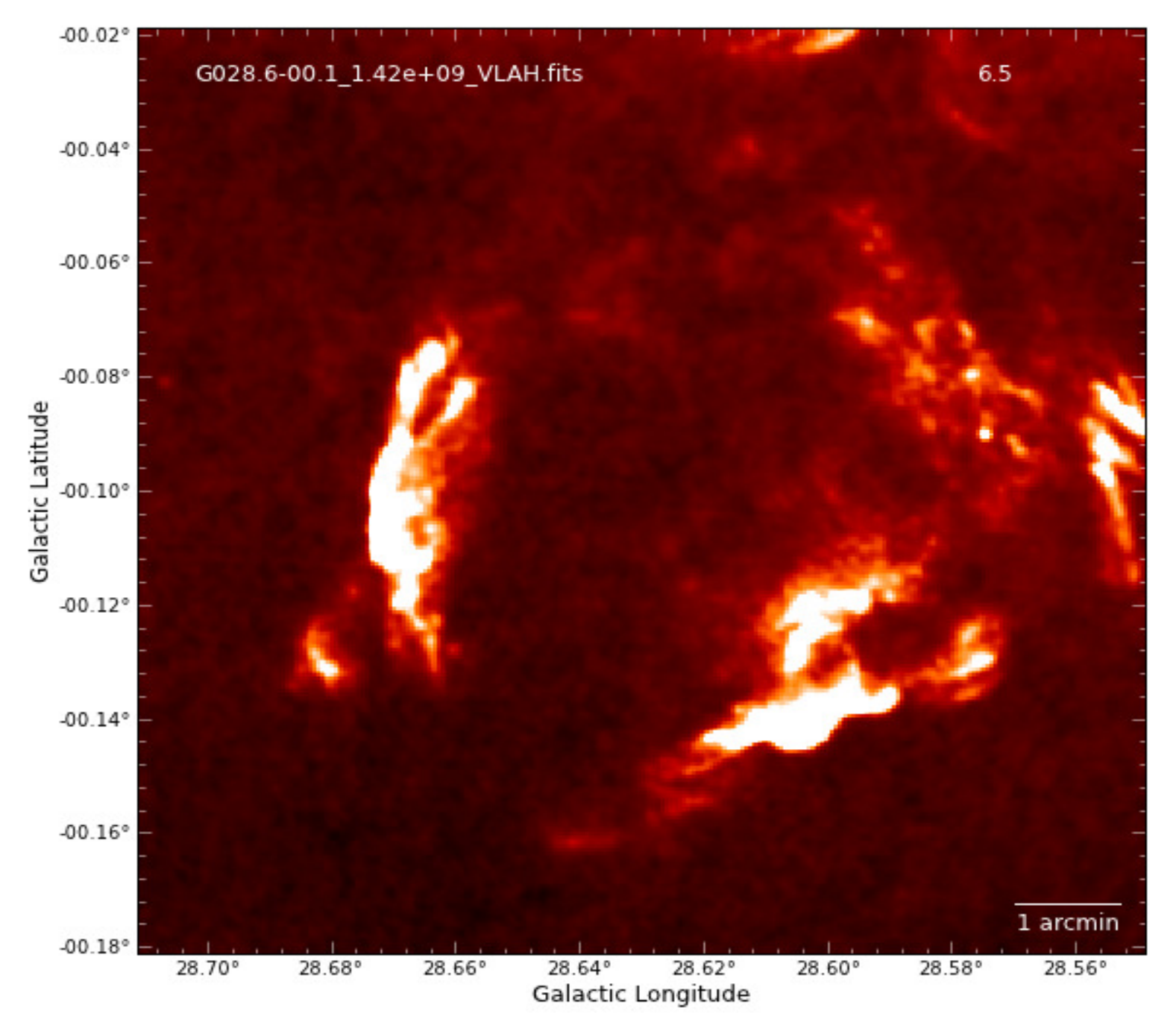}
    \caption{The upper three images show the simulation relative radio flux density at different ages.
    The lower images show the observed radio images of SNRs, G53.6-2.2, G29.7-0.3 and G28.6-0.1 \citep{West2016},
    all of which are bilateral asymmetric.}
\label{fig:parc}
\end{figure*}

Figure~\ref{fig:par} shows that the radio morphology is a bilateral asymmetric SNR.
\citet{vanMarle2014} showed that the magnetic field would shape the stellar wind nebulae of asymptotic giant (AGB) stars
as bilateral symmetric morphologies.
Including the motion of ABG stars,  \citet{vanMarle2014a} studied the instabilities in such a system.
\citet{Meyer2017} also simulated the bow shock nebulae of hot massive stars in a magnetized medium, which shows
similar results as our parallel stellar wind simulation.
However, they did not add the supernova explosion and convert the results to radio images.
Taking the circular SNR into account, we are able to simulate five types of SNRs in our classification.
Only the multi-layers and the irregular SNRs are difficult to be simulated.
Their formations are likely influenced by the inhomogeneous initial surrounding environment or the unusual progenitor
\citep{Orlando2007,Orlando2017}.

The upper images of Figure~\ref{fig:parc} show the simulation morphologies at 1450, 1850 and 3050 years respectively.
As a comparison, three real SNRs, G53.6-2.2, G29.7-0.3, G28.6-0.1, are shown in the lower panels of Figure~\ref{fig:parc}.
Because of the similar morphologies between the simulation images and the observation images, the three SNRs are likely all
few thousands years old.
In fact, G29.7-0.3 is about one thousands years old \citep{Leahy2008} and G28.6-0.1 \citep{Bamba2001} is no more than
2700 years old.
G53.6-2.2 seems older (about 15,000 years old, see \citet{Long1991}), which is worthy to be further checked.
In addition, the X-ray emissions of the three SNRs are all more or less separated from the radio shell \citep{Broersen2015,Su2009,Bamba2001},
similar to SNR G116.9+0.2.
The simulation results also coincide with these observations, just like the perpendicular simulation for G116.9+0.2，
so we do not show them here.

Since the parameters are same at the two simulations, we are able to compare the relative flux density in parallel with that in perpendicular
simulations at same age.
Figure~\ref{fig:par} shows the relative flux density in the y-z plane for the parallel simulation is much lower than that
for the perpendicular simulation.
In other words, bilateral asymmetric SNRs should be less than unilateral small-radian SNRs.
This is supported by the statistics in Table.~\ref{table:stat}.
The unilateral large-radian SNRs should be less than the bilateral asymmetric SNRs, if we only take the x-z plane into consideration in the simulation
results.
However, the directions of the LoS might influence this estimation.
For example, the Figure~\ref{fig:45deg} shows a unilateral large-radian SNR is brighter than the bilateral asymmetric SNR.
In fact, Table.~\ref{table:stat} implies that the unilateral large-radian SNRs are more than the bilateral asymmetric SNRs.

\section{Summary}
Taking the evolution result of the stellar wind as the initial conditions, we simulate the SNR evolution of a runaway 40 M$_{\odot}$ progenitor star.
The stellar wind simulations includes two models, the perpendicular simulation and the parallel simulation.
Based on real radio morphologies, we classify the SNRs into seven types.
Our conclusions are summarized as follows:
\begin{enumerate}

    \item The stellar wind of the massive progenitor plays a key role in shaping the radio morphologies of SNRs, and is possibly important more than the
    initial surrounding environment.

    \item Considering the stellar wind, we can explain many radio morphologies of SNRs, except for the multi-layers and irregular SNRs.

    \item It is not suggested to infer the large-scale magnetic field or density distribution in Milky Way based on the radio morphologies of SNRs.

    \item The thermal conduction might slightly influence the SNR radio morphologies, but is not very important.

    \item The separation between X-ray and radio emission of some SNRs is possibly related with the motion of the progenitor.

\end{enumerate}
We note that there are many simplifications in our current work.
It will be interesting to study the formation of multi-layers and irregular SNRs by more detailed simulation in the near future, e.g. including
an inhomogeneous initial surrounding environment or a special progenitor, etc.

\acknowledgements
We thank Dr.Meyer for his explaining the thermal conduction of the stellar wind.
We acknowledge support from the NSFC (11473038).

\software{PLUTO \citep{Mignone2007,Mignone2012}}

\bibliographystyle{aasjournal}
\bibliography{./mydb}
\end{document}